\documentclass[Afour,times,sageh]{sagej}
\usepackage{natbib}
\usepackage{moreverb,url}
\usepackage[colorlinks,bookmarksopen,bookmarksnumbered,citecolor=red,urlcolor=red]{hyperref}
\newcommand\BibTeX{{\rmfamily B\kern-.05em \textsc{i\kern-.025em b}\kern-.08em
T\kern-.1667em\lower.7ex\hbox{E}\kern-.125emX}}
\def\volumeyear{2022}
\usepackage{tikz,xcolor,hyperref}

\definecolor{lime}{HTML}{A6CE39}
\DeclareRobustCommand{\orcidicon}{%
	\begin{tikzpicture}
	\draw[lime, fill=lime] (0,0) 
	circle [radius=0.16] 
	node[white] {{\fontfamily{qag}\selectfont \tiny ID}};
	\draw[white, fill=white] (-0.0625,0.095) 
	circle [radius=0.007];
	\end{tikzpicture}
	\hspace{-2mm}
}

\foreach \x in {A, ..., Z}{%
	\expandafter\xdef\csname orcid\x\endcsname{\noexpand\href{https://orcid.org/\csname orcidauthor\x\endcsname}{\noexpand\orcidicon}}
}

\begin{document}

\runninghead{Kristiansen \textit{et~al.}}

\title{The Visual Survey Group: A Decade of Hunting Exoplanets and Unusual Stellar Events with Space-Based Telescopes}

\author{Martti H. K. Kristiansen\affilnum{1}\orcidA{}, Saul A. Rappaport\affilnum{2}\orcidB{}, Andrew M. Vanderburg\affilnum{2}\orcidC{}, Thomas L. Jacobs\affilnum{3}\orcidD{}, Hans Martin Schwengeler\affilnum{4}\orcidE{}, Robert Gagliano\affilnum{5}\orcidF{}, Ivan A. Terentev\affilnum{6}\orcidG{}, Daryll M. LaCourse\affilnum{7}\orcidH{}, Mark R. Omohundro\affilnum{4}, Allan R. Schmitt\affilnum{8}\orcidL{}, Brian P. Powell\affilnum{9}\orcidJ{} and Veselin B. Kostov\affilnum{9,10,11}\orcidK{}}

\affiliation{\affilnum{1}Brorfelde Observatory, Observator Gyldenkernes Vej 7, DK-4340 Tølløse, Denmark\\
\affilnum{2}Department of Physics, and Kavli Institute for Astrophysics and Space Research, M.I.T., Cambridge, MA 02139, USA\\
\affilnum{3}Amateur Astronomer, 12812 SE 69th Place, Bellevue, WA 98006\\
\affilnum{4}Citizen scientist, c/o Zooniverse, Department of Physics, University of Oxford, Denys Wilkinson Building, Keble Road, Oxford, OX1 3RH, UK\\
\affilnum{5}Amateur Astronomer, Glendale, Arizona\\
\affilnum{6}Citizen Scientist, Petrozavodsk, Russia\\
\affilnum{7}Amateur Astronomer, 7507 52nd Pl NE, Marysville, WA 98270\\
\affilnum{8}Citizen Scientist, 616 W. 53rd. St., Apt. 101, Minneapolis, MN 55419, USA\\
\affilnum{9}NASA Goddard Space Flight Center, 8800 Greenbelt Road, Greenbelt, MD 20771, USA\\
\affilnum{10}SETI Institute, 189 Bernardo Ave, Suite 200, Mountain View, CA 94043, USA\\
\affilnum{11}GSFC Sellers Exoplanet Environments Collaboration}

\corrauth{Martti H. Kristiansen, 
Brorfelde Observatory, 
Observator Gyldenkernes Vej 7, 
DK-4340 Tølløse, Denmark}
\email{martti@outinto.space \& marki@holb.dk}

\begin{abstract}
This article presents the history of the Visual Survey Group (VSG) - a Professional-Amateur (Pro-Am) collaboration within the field of astronomy working on data from several space missions (Kepler, K2 and TESS). This paper covers the formation of the VSG, its survey-methods including the most common tools used and its discoveries made over the past decade. So far, the group has visually surveyed nearly 10 million light curves and authored 69 peer-reviewed papers which mainly focus on exoplanets and discoveries involving multistellar systems found using the transit method. The preferred manual search-method carried out by the VSG has revealed its strength by detecting numerous sub-stellar objects which were overlooked or discarded by automated search programs, uncovering some of the most rare stars in our galaxy, and leading to several serendipitous discoveries of unprecedented astrophysical phenomena. The main purpose of the VSG is to assist in the exploration of our local Universe, and we therefore advocate continued crowd-sourced examination of time-domain data sets, and invite other research teams to reach out in order to establish collaborating projects.
\end{abstract}

\keywords{Pro-Am, amateur astronomy, citizen science, exoplanetary systems, eclipsing binaries, dipper stars, black swans}

\maketitle

\begin{figure*}
\setlength{\fboxsep}{0pt}%
\setlength{\fboxrule}{0pt}%
\begin{center}
\includegraphics[width=17cm]{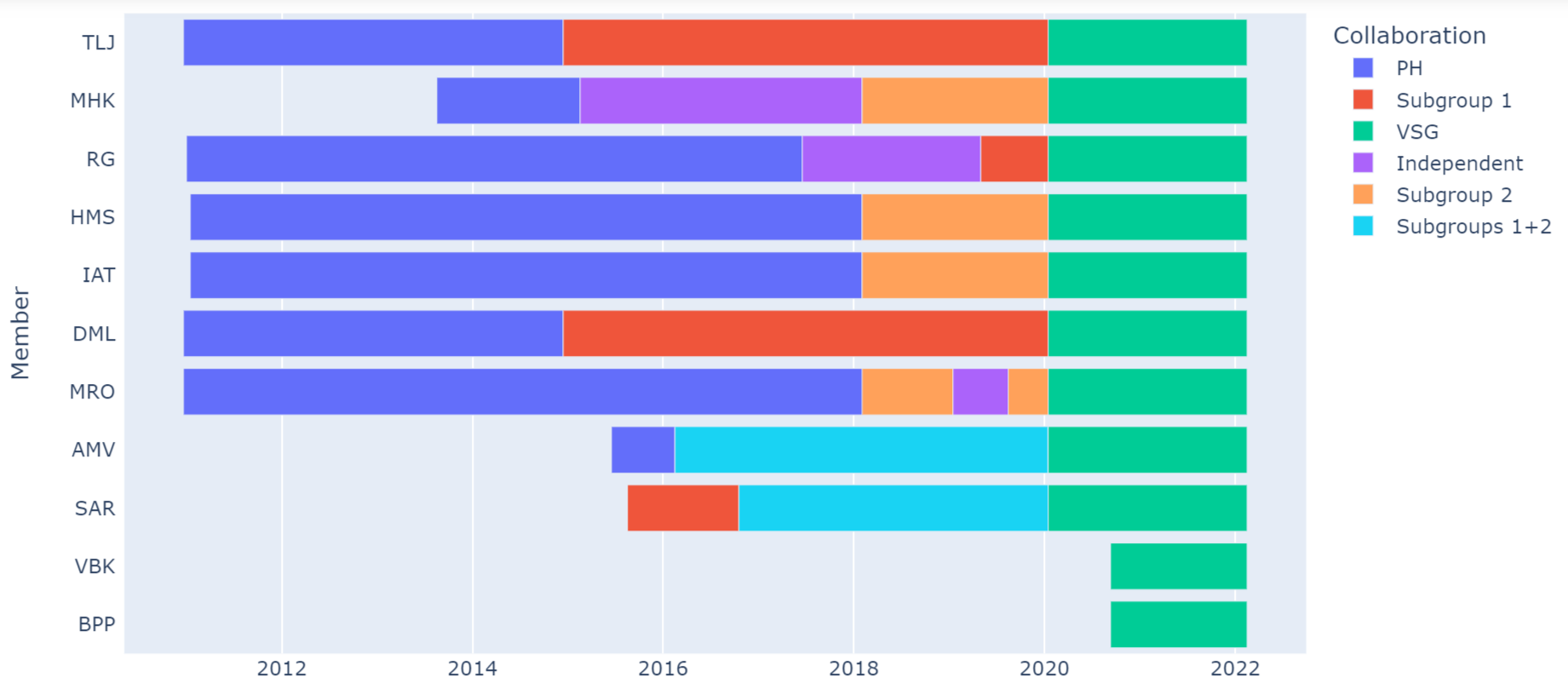}
\end{center}
\caption{Timeline of the individual members. The different collaborations include Planet Hunters (PH, blue), Subgroup 1 and 2 (red, orange and cyan), independent activities (purple) all leading towards the VSG-collaboration (VSG, green).
\label{F1}}
\end{figure*}

\section{Introduction}
The field of astronomy has a long history of collaboration between professional and amateur astronomers dating back centuries (see e.g. \cite{boyd2011pro}, \cite{lintott2020citizen}). In recent times, largescale projects such as MilkyWay@Home \citep{newberg2013milkyway}, SETI@home \citep{anderson2000seti}, Stardust@home \citep{westphal2005stardust} and Galaxy Zoo \citep{lintott2008galaxy} became widespread, counting millions of participant volunteers to date. The popularity of Galaxy Zoo evolved into the citizen science platform Zooniverse which at the time of writing has 2.455.914 registered users alone\endnote{\href{https://www.zooniverse.org/}{https://www.zooniverse.org/}}. One particular project from this platform, Planet Hunters (PH, \cite{fischer2012planet}),  was created on 16 December 2010 alongside the launch of the \textit{Kepler} spacecraft and its wonderful photometric data stream \citep{koch2010kepler}. In the PH-project, citizen scientists were given an optional tutorial and shown light curves in which they were asked to identify exoplanetary transits \citep{lintott2013planet}. This effort was intended to serve as an aid to the Transiting Planet Search algorithm developed by the Kepler team (TPS, \cite{jenkins2010overview}, \cite{jenkins2010transiting}). Initially, in order for a suspected Planet Candidate (PC) to be upgraded to a Threshold Crossing Event (TCE), the TPS-algorithm needed at least three consistent transits with a statistical significance $\geq 7.1 \sigma$ \citep{jenkins2002some}. Consequently, these criteria left a large portion of parameter space initially unexplored, but at the same time, an open door for focused visual surveying by citizen scientists. Those collective efforts quickly proved to be particularly effective with respect to the dearth of long-period exoplanet detections, the improvement of which betters the census of planet occurrence rates\endnote{\href{https://blog.planethunters.org/2014/09/11/planet-occurrence-rates-2/}{Planet Occurrence Rates}}.
In the primary Kepler mission \citep{borucki2010kepler}, few long-period exoplanets (i.e. planets with $<$ 3 transits) were identified by professional astronomers (see e.g. \cite{batalha2013planetary}, \cite{foreman2016population}, \cite{uehara2016transiting}, \cite{schmitt2017search}). Similar circumstances emerged for the \textit{K2}-mission \citep{howell2014k2}. Although the presence of more single transits was expected for K2, due to its 80-day observation campaigns, few were detected (see e.g. \cite{osborn2016single}). In the Transiting Exoplanet Survey Satellite mission (\textit{TESS}, \cite{ricker2014transiting}), the estimated single transit harvest exceeds that of K2 (\cite{cooke2018single}, \cite{villanueva2019estimate}), partly due to even shorter observation intervals ($\approx$ 27 days).\\
In this paper, we describe the efforts of a Professional-Amateur (Pro-Am) collaboration called the Visual Survey Group (VSG) with the goal of searching for long-period planets and other unusual astrophysical phenomena. Even though astronomers have developed new methods (e.g. \cite{olmschenk2021identifying}; \cite{osbornmonotools}; \cite{cui2021identify}), the hunt for single transits is time consuming and incomplete. Visual surveying therefore continues to be a viable detection method for these signals. Moreover, some of the new automated methods designed to detect irregular light curve features (The Weird Detector, \cite{wheeler2019weird}) and exocomets \citep{kennedy2019automated} were inspired by discoveries made by the VSG. Therefore, in spite of significant advances in automated search approaches over the last decade, the pattern recognition ability of the human mind and the steadfast participation of amateurs is far from being rendered a useless or frivolous endeavor.\\
In this work, we describe the history of the VSG (Section 2), while Section 3 covers the manual and visual search approach primarily undertaken by the VSG and tools used. Section 4 discusses the scientific discoveries made by the VSG which amongst others consist of exoplanets, light curve anomalies which have turned out to be all manner of dusty occultations, including exocomets, and eclipsing binaries which we have found to be part of triply eclipsing triples, and higher-order stellar systems. In Section 5 we carry out a discussion and our conclusion.

\section{The Visual Survey Group}
The Visual Survey Group (hereafter, VSG) consists of seven citizen scientists (TLJ, RG, MRO, DML, IAT, HMS and MHK) from four countries (USA, Russia, Switzerland and Denmark). Their professions (or former employment in the case of retirement) range over the fields of business, medicine, programming, aerospace engineering, information technology, mathematics and astronomy. Three group members are formally retired. In addition, two professional astronomers situated at the Massachusetts Institute of Technology (MIT; SAR and AV) and two at NASA's Goddard Space Flight Center (NASA GSFC; BPP and VBK) complete the VSG-collaboration.\\ 
The establishment of the VSG is outlined in the following subsection but for the purpose of historical recordkeeping, we here mention three individuals who no longer are affiliated with the VSG but nonetheless made important contributions in the earlier stages: Troy Winarski, Alexander Venner and Kian J. Jek (see references in Sect. 4 and \cite{venner2021true}). Kian J. Jek received the American Astronomical Society Chambliss Amateur Achievement Award in 2012 which also was awarded to a VSG-member (DML) in 2016\endnote{\href{https://aas.org/grants-and-prizes/chambliss-amateur-achievement-award}{The AAS's Chambliss Amateur Achievement Award recipients}}.

\begin{figure*}
\setlength{\fboxsep}{0pt}%
\setlength{\fboxrule}{0pt}%
\begin{center}
\includegraphics[width=17cm]{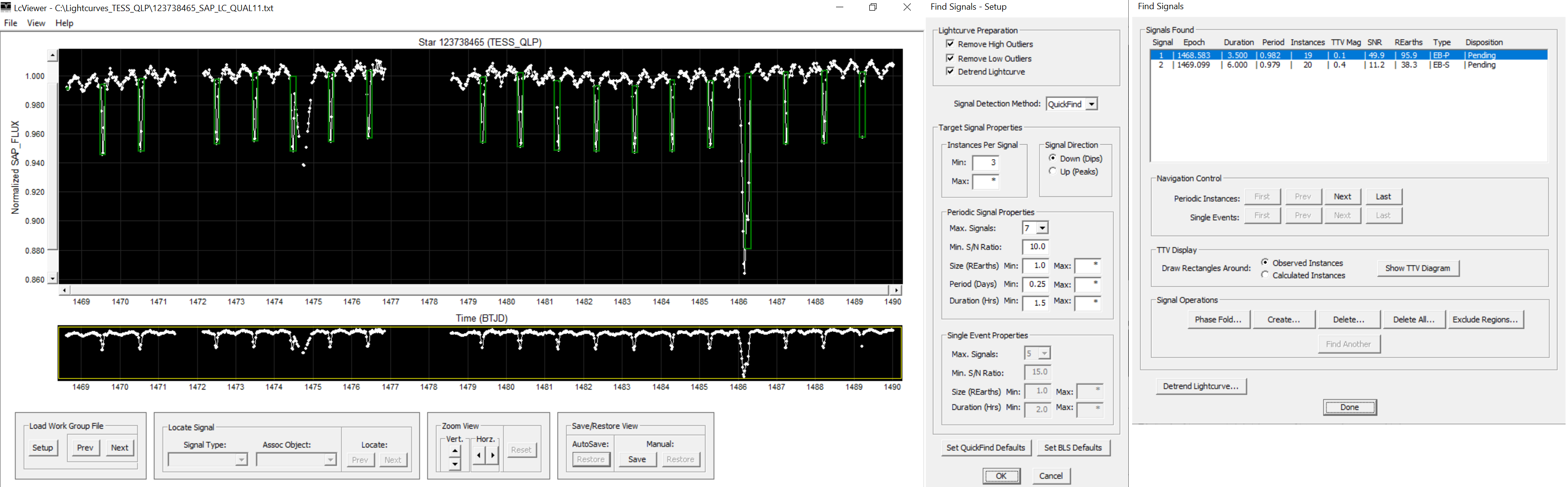}
\end{center}
\caption{A TESS Quick Look Pipeline (QLP) light curve for TIC 123738465 showing a multistellar candidate signal. Highlighted in green, is the primary signal of the short-period eclipsing binary detected by LcViewer via its automated QuickFind search feature.
\label{F2}}
\end{figure*}

\subsection{Common Grounds}
All members met on the Zooniverse platform at the now defunct, original Planet Hunters project led by Yale University between 2010 and 2013. For each individual, the exact pathway towards the VSG-collaboration varies and is therefore summarized in a timeline in Figure~\ref{F1}. At the outset, the hunt for exoplanets was done using ingested data from the Kepler mission \citep{borucki2010kepler}, which were shown in light curve snips (light curves from single Kepler-Quarters were divided in smaller portions) followed by the question: ``Does the star have any transit features?" (Fig. 1, \cite{lintott2013planet}). Subsequently, the users could either continue to the next classification or get involved in a discussion concerning the most recently classified star (`Talk-section'). The Talk-section connected the VSG-members and was also heavily used by other so-called `superusers' (see Sect. 2 in \cite{schwamb2013planet}).\\
By the time of Kepler's repurposed two-wheel mission, K2 \citep{howell2014k2}, the PH-website was redesigned. It was during the K2-revival that AV joined PH as an official Science Team member and TLJ, DML and MHK made contact with SAR on a more informal basis. All seven citizen scientists made the transition to the new PH-website, however a lack of certain improvements of the interface backend, specifically common to how potential targets of interest were tracked, tallied and subsequently analyzed, resulted in this being a limited visit for some individuals (first TLJ, DML and later MHK). This said, connections to the PH-science team were maintained, and some of the VSG-members were engaged in public outreach (see e.g. the PH-science team's Reddit science `Ask me anything'\endnote{\href{https://www.reddit.com/r/science/comments/3ebavu/science_ama_series_were_the_planet_hunters_team/}{Science AMA Series: We’re the Planet Hunters team using crowd-sourcing to search for exoplanets in the Kepler space telescope data. Ask us anything (and join the search)!}}, PH-interviews\endnote{\href{https://blog.planethunters.org/2018/01/07/without-planet-hunters-none-of-the-subsequent-discoveries-would-have-been-possible/}{“Without Planet Hunters, none of the subsequent discoveries would have been possible.”}} and earlier PH-blog posts\endnote{\href{https://blog.planethunters.org/2011/12/}{PH-blog posts by TLJ, DML and Kian J. Jek.}}).\\
Meanwhile, a new Zooniverse project arose which also dealt with data from K2, the Exoplanet Explorers\endnote{\href{https://www.zooniverse.org/projects/ianc2/exoplanet-explorers}{https://www.zooniverse.org/projects/ianc2/exoplanet-explorers/}} \citep{christiansen2018k2}, and later a third project, Planet Hunters TESS\endnote{\href{https://www.zooniverse.org/projects/nora-dot-eisner/planet-hunters-tess}{https://www.zooniverse.org/projects/nora-dot-eisner/planet-hunters-tess}} \citep{eisner2020planet} working on TESS-data. From the VSG, HMS and IAT participated in both projects.\\
Instead of running several parallel surveys of similar nature, the subteams joined forces in January 2020 and once again found common ground. In doing so, the workflow was streamlined in both light curve surveying and vetting. Two years later, in January 2022, BPP and VBK formally joined the VSG, adding two more professional colleagues to the collaboration which had begun in September 2020.

\begin{figure*}
\setlength{\fboxsep}{0pt}%
\setlength{\fboxrule}{0pt}%
\begin{center}
\includegraphics[width=17cm]{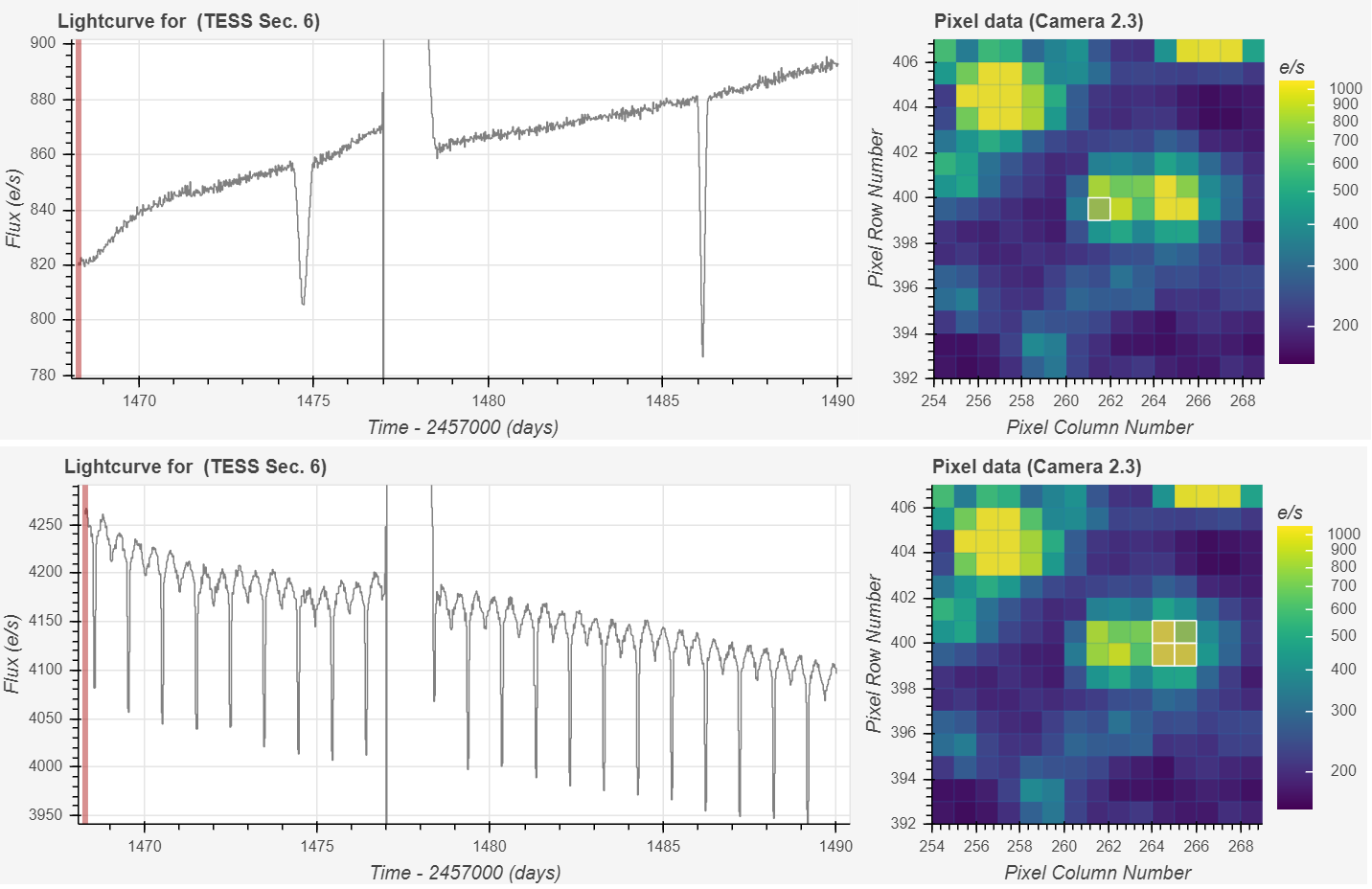}
\end{center}
\caption{Target Pixel File analysis of TIC 123738465 using the interactive graphical user interface from the Lightkurve software showing that the multistellar candidate from Figure 2 is a false positive. The figure shows two light curves (left), for two stars with two different photometric aperture sizes (right) indicated with white bordered regions (1x1 pixel and 2x2 pixels, respectively).
\label{F3}}
\end{figure*}

\section{Approach, Tools and Data} 
The VSG operates on Linux, Macintosh, and Windows systems. The primary approach has since the beginning been visual and manual surveying, which still has many advantages compared to state-of-the-art computer software. All communications about potential discoveries are made within a day via e-mail contact to all members. When members find potentially interesting new objects they are immediately shared among the entire VSG via e-mail. Discussions about the viability of each discovery is then continued via a continuous e-mail stream.

\subsection{The LcTools Approach}
Since 2012, the VSG-members have worked closely with the \texttt{LcTools} project (\cite{kipping2015hunt},  \cite{schmitt2019lctools}, \cite{schmitt2021lctools}) whose purpose is to provide citizen scientists, students, and universities worldwide with custom applications, data, and consulting services for viewing and analyzing light curves from NASA space missions. Based on their research needs and objectives, the VSG members have contributed many ideas for improvement in the product and have assisted in testing the product prior to public release.\\ \texttt{LcTools} has revolutionized visual surveying of stellar light curves. Since becoming available in 2012, the product has been gradually upgraded with sophisticated packages such as \texttt{VARTOOLS} \citep{hartman2012vartools}. However, its manual and visual features should undoubtedly be credited with the underlying success of the VSG. This Windows-based software allows its users to efficiently scan light curves in a matter of seconds. For instance, upon release of a K2-campaign consisting of $\approx$ 20,000 light curves, a VSG-member could normally survey the entire data set within a day.\\
\texttt{LcTools} provides many user-friendly custom-designed features, some not found in any other publicly available product. These include, but not limited to (1) a near instantaneous display of light curves using a simple one-button click to rapidly scan through a large list of light curve files in a directory, (2) optimized light curve presentation to facilitate the visual detection of signals, (3) fast and efficient light curve navigation operations such as panning and zooming, (4) real-time tracking of time and flux at the cursor, (5) ability to measure signals and time/flux intervals using the mouse, (6) display of mission-based and community-based signals such as TCEs, Kepler Objects of Interest (KOIs), K2 Objects of Interest (K2OIs), Community TESS Objects of Interest (CTOIs) and TESS Objects of Interest (TOIs) to avoid known signals being re-discovered, (7) automatic detection of periodic signals and single events using the QuickFind and Box Least Squares (BLS) signal detection methods, (8) recording and display of user defined signals, (9) ability to phase-fold periodic signals over different time domains, (10) manual detrending of light curves based on easy-to-use flattening levels, (11) display of the host star properties for a light curve, (12) creation of signal based property reports in Excel, (13) building of light curve files (individually or bulk) for the Kepler, K2 and TESS missions and associated High Level Science Products (HLSPs) using the source data from the Mikulski Archive for Space Telescopes (MAST)\endnote{\href{https://archive.stsci.edu/prepds/index.html/}{https://archive.stsci.edu/prepds/index.html/}}.\\
At present, light curve files are obtained in a .txt-format from the \texttt{LcTools} website\endnote{\href{https://sites.google.com/a/lctools.net/lctools/}{https://sites.google.com/a/lctools.net/lctools/}} if available or built using the LcGenerator application in \texttt{LcTools}. Otherwise, data are directly obtained at the source (see Tab. 1).
Most of the discoveries made by the VSG (Sect. 4) were initially flagged using the LcViewer application in \texttt{LcTools}. Figure~\ref{F2} shows the Quick Look Pipeline (QLP) light curve for TIC 123738465 in LcViewer and the accompanying user interface for finding signals automatically using the QuickFind signal detection method.\\

\subsection{Additional Survey Tools}
During the \textit{Kepler} mission \citep{borucki2010kepler}, and prior to the first release of \texttt{LcTools}, the seven citizen scientists primarily made use of the PH-interface for light curve classification and the MAST for data acquisition. This was accompanied by services such as Tool for OPerations on Catalogues And Tables (\texttt{TOPCAT}, \cite{taylor2011topcat}) and \texttt{Fv} \citep{pence2012fv} used for light curve files and Target Pixel Files (TPFs), the SkyView Query Form\endnote{\href{https://skyview.gsfc.nasa.gov/current/cgi/query.pl}{https://skyview.gsfc.nasa.gov/current/cgi/query.pl}} to identify the stellar vicinity of a target star, and the Amateur Kepler Observer (\texttt{AKO}, Winarski (private software)) to search for Transit Timing Variations (TTVs), i.e. in order to produce `Winarski-plots' which now are commonly known as river-plots (see e.g. Fig. 4 in \cite{agol2018transit}).\\
In the K2-era \citep{howell2014k2}, VSG also made use of \texttt{VESPA} to statistically validate transiting exoplanets \citep{morton2015vespa}, and \texttt{Kadenza} \citep{barentsen2018kadenza} in order to obtain raw cadence pixel files for quick-views of targets of interest, and other light curve extraction software for operating TPFs (\texttt{PyKE}, \cite{still2012pyke}; \texttt{AKO-TPF}, (Winarski, private software)). The latter is comparable to the interactive features of \texttt{Lightkurve} \citep{cardoso2018lightkurve}, which now is a standard ingredient in the VSG-vetting process for TESS data. In the VSG, \texttt{Lightkurve} is mostly used to look for contamination, e.g. EBs mimicking PCs, and Solar System Objects (SSOs) mimicking PCs or stellar flares/outbursts. These SSOs are identified using the Sky Body Tracker (\texttt{SkyBoT}, \cite{berthier2006skybot}, \cite{berthier2016prediction}). Furthermore, \texttt{Lightkurve} has been proven extremely useful when assessing hierarchical eclipsing candidates due to the large pixel size (21$''$) of \textit{TESS} as illustrated in Figure~\ref{F3}. During the TESS-mission, TPFs have also been obtained using \texttt{TESScut}\endnote{\href{https://mast.stsci.edu/tesscut/}{https://mast.stsci.edu/tesscut/}} \citep{brasseur2019astrocut}.\\
Information concerning stellar parameters can be found at the Exoplanet Follow-up Observing Program (ExoFOP)\endnote{\href{https://exofop.ipac.caltech.edu/}{https://exofop.ipac.caltech.edu/t}}, the Gaia-collaboration\endnote{\href{https://gea.esac.esa.int/archive/}{https://gea.esac.esa.int/archive/}} (\cite{collaboration2016gaia}, \cite{brown2021gaia}), the Aladin Lite finding charts at Swarthmore%\endnote{\href{https://astro.swarthmore.edu/transits/aladin_finder.html}{https://astro.swarthmore.edu/transits/aladin_finder.html}}
, the MAST or at databases operated by the Strasbourg astronomical Data Center (CDS)\endnote{\href{http://cdsweb.u-strasbg.fr/about}{http://cdsweb.u-strasbg.fr/about}} - SIMBAD \citep{wenger2000simbad} and ViZieR \citep{genova2000cds}. In addition, while evaluating dipper candidates (Sect. 4.4), the online search engine\endnote{\href{https://irsa.ipac.caltech.edu/applications/wise/}{https://irsa.ipac.caltech.edu/applications/wise/}} of the Wide-field Infrared Survey Explorer (WISE, \cite{wright2010wide}) is used to search for IR-excess. Moreover, custom programs written in python, C, Fortran, and JavaScript languages have been created by the VSG over the years.\\
In addition to this work, the analysis, expertise and guidance from the members at MIT and NASA help ensure the prospects of follow-up observations at suitable facilities. Follow-up also includes searching for archival photometry at the Digital Access to a Sky Century @ Harvard (DASCH, \cite{grindlay2009dasch}), the All Sky Automated Survey (ASAS, \cite{pojmanski1997all2}), the All Sky Automated Survey for Supernovae (ASAS-SN, \cite{shappee2014all}) and the Asteroid Terrestrial-impact Last Alert System (ATLAS, \cite{tonry2018atlas}).\\
When validating a target, several steps are taken into consideration: For each source that looks potentially interesting, the VSG-team, usually starting with the person who initially found the source, checks such archival resources as (i) SIMBAD, (ii) the WISE-images in four bands, (iii) PanSTARRS-images \citep{chambers2016pan}, (iv) the VizieR spectral energy distribution and (v) Gaia.  These provide some immediate indication of (i) what is already known about the object and pointers to the literature, (ii) whether the object has any obvious near infrared excess, (iii) if there are any nearby neighbors that might be contaminating the signal, (iv) what is known about the overall spectral shape of the source, and (v) to see if the star has any bound neighbors. This individual also checks for earlier {\it TESS} sectors in which the source might have been observed. Archival data from such publicly available resources as ASAS-SN, ATLAS, and DASCH are downloaded to investigate the source activity over intervals of six years to possibly a century. We also check with team members of ground-based surveys to determine if there is any archival data for these sources (see e.g. \cite{rappaport2022six} and references herein).\\
There are a few members of the VSG-team who are expert in measuring the light centroid of the time varying part of the signal, and this is checked to be certain that the correct star has been identified. In cases where potential planet candidates have been found, initial checks of the folded light curve are made to inspect transit shapes and to check for odd-even effects. Our group also has access to many experts in various subfields of astronomy with whom we consult for opinions about classes of objects that are outside the immediate expertise of the professional astronomy members of the VSG-team. If necessary, we have access to astronomers who can take spectra or high-resolution images for us (speckle or adaptive optics).

\subsection{Data under the Bridge}

Table~\ref{T1} lists different data sources including numbers of light curves manually surveyed by the VSG. The estimated numbers do not refer to unique stars due to overlaps for the three missions, and for K2 and TESS, also overlapping campaign and sector targets. In cases where a light curve has been surveyed multiple times, or by additional surveyors, it only counts once. In this manner, the VSG has so far surveyed nearly 10 million distinct light curves. 

\begin{table}[h]
\small\sf\centering
\caption{Overview of data sources and the number of light curves manually surveyed by the VSG. Entries do not correspond to unique stars due to overlaps between data sources and missions.\label{T1}}
\begin{tabular}{lll}
\toprule
Data source&Quarter/Campaign/Sector&Light curves\\
\midrule
Kepler &Q1 - 17 &181,300\\
K2 SFF &ET - C19 &421,600\\
TIC CTL &S1 - 48 &907,200\\
QLP &S2-3, S13, S17-21, S16-27 &5,914,000\\
OELKER &S1 - 5 &543,300\\
CDIPS &S6 - 13 &67,200\\
PATHOS &S4 - 14 &31,800\\
SPOC &S22, S30 &31,200\\
TICA &S35 &2,200\\
GSFC &S1 - 40 &1,573,000\\
\midrule
Total & &9,672,800\\
\bottomrule
\end{tabular}\\[10pt]
\end{table}

On the PH-website, it is impossible to assess the number of Kepler-targets \citep{brown2011kepler} surveyed by the individual VSG-members. However, it is fair to estimate that all seven citizen scientists have surveyed more than half of the stars in the primary Kepler mission, and the entire Kepler-data set was later fully scrutinized by TLJ twice using \texttt{LcTools}. All seven citizen scientists have individually surveyed the entire K2 Self Flat Fielding (SFF) data set from \cite{vanderburg2014technique}.\\
In the TESS-mission a number of data sources have been surveyed including the TESS Input Catalog Candidate Target List (TIC CTL, \cite{stassun2018tess}), the Science Processing Operations Center (SPOC, \cite{caldwell2020tess}), OELKER \citep{oelkers2018precision}, the PSF-based Approach to TESS High quality data Of Stellar
clusters (PATHOS, \cite{nardiello2019psf}), the Cluster Difference Imaging Photometric Survey (CDIPS, \cite{bouma2019cluster}), the MIT Quick Look Pipeline (QLP, \cite{huang2020photometry1}, \cite{huang2020photometry2}), the TESS Image CAlibrator Full Frame Images (TICA, \cite{fausnaugh2020calibrated}) from which we generate custom light curves of one-orbit TESS-data previews that are available earlier than the standard mission products, and the Goddard Space Flight Center (GSFC, see Sect. 2, \cite{powell2021tic}). For a combined, more detailed description of the various TESS data sources, we direct the reader to Section 2.1 of Capistrant et al. (in-review).

\begin{figure*}
\begin{center}
\includegraphics[width=0.42\textwidth]{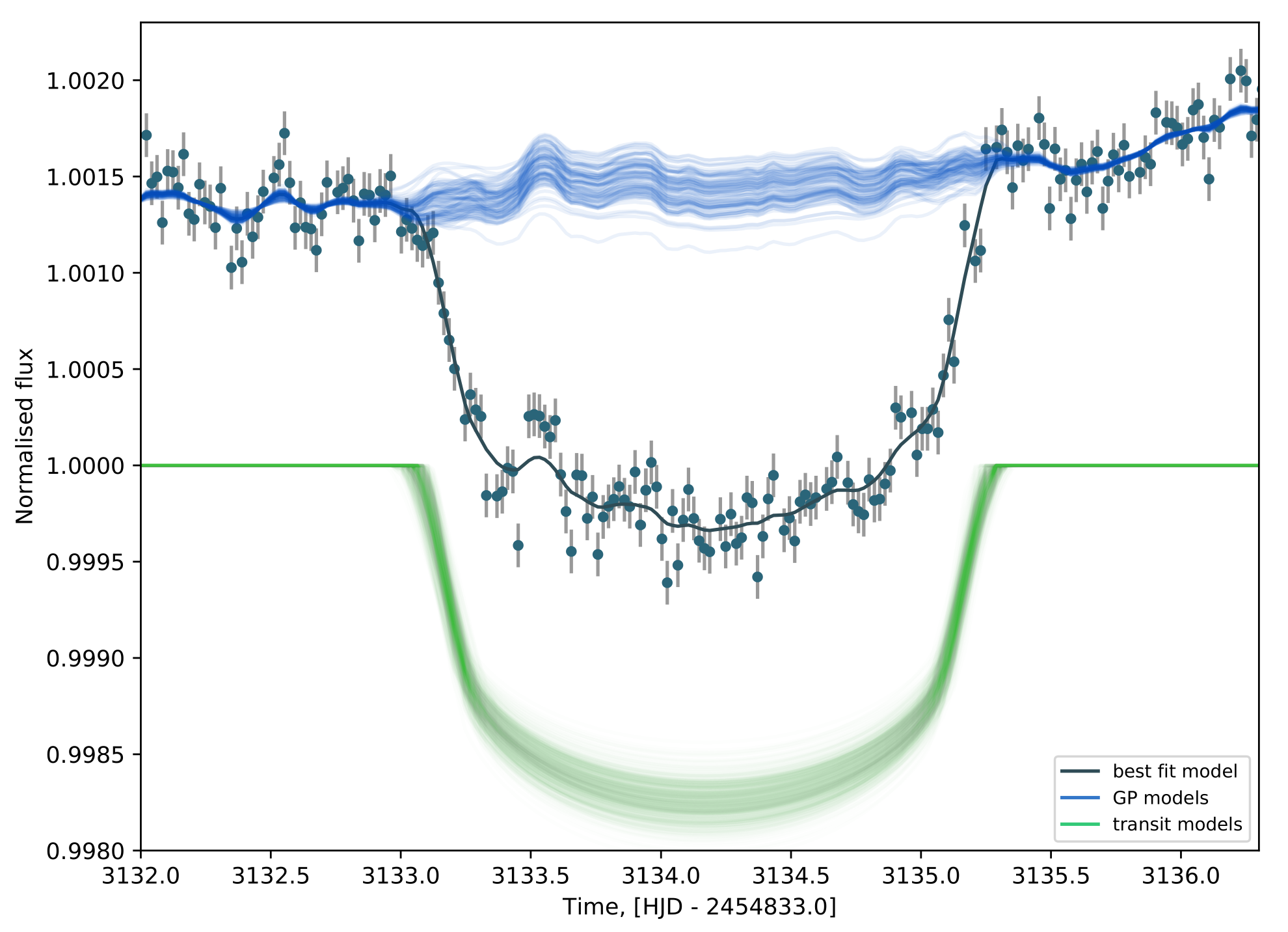} \hglue0.4cm
\includegraphics[width=0.44\textwidth]{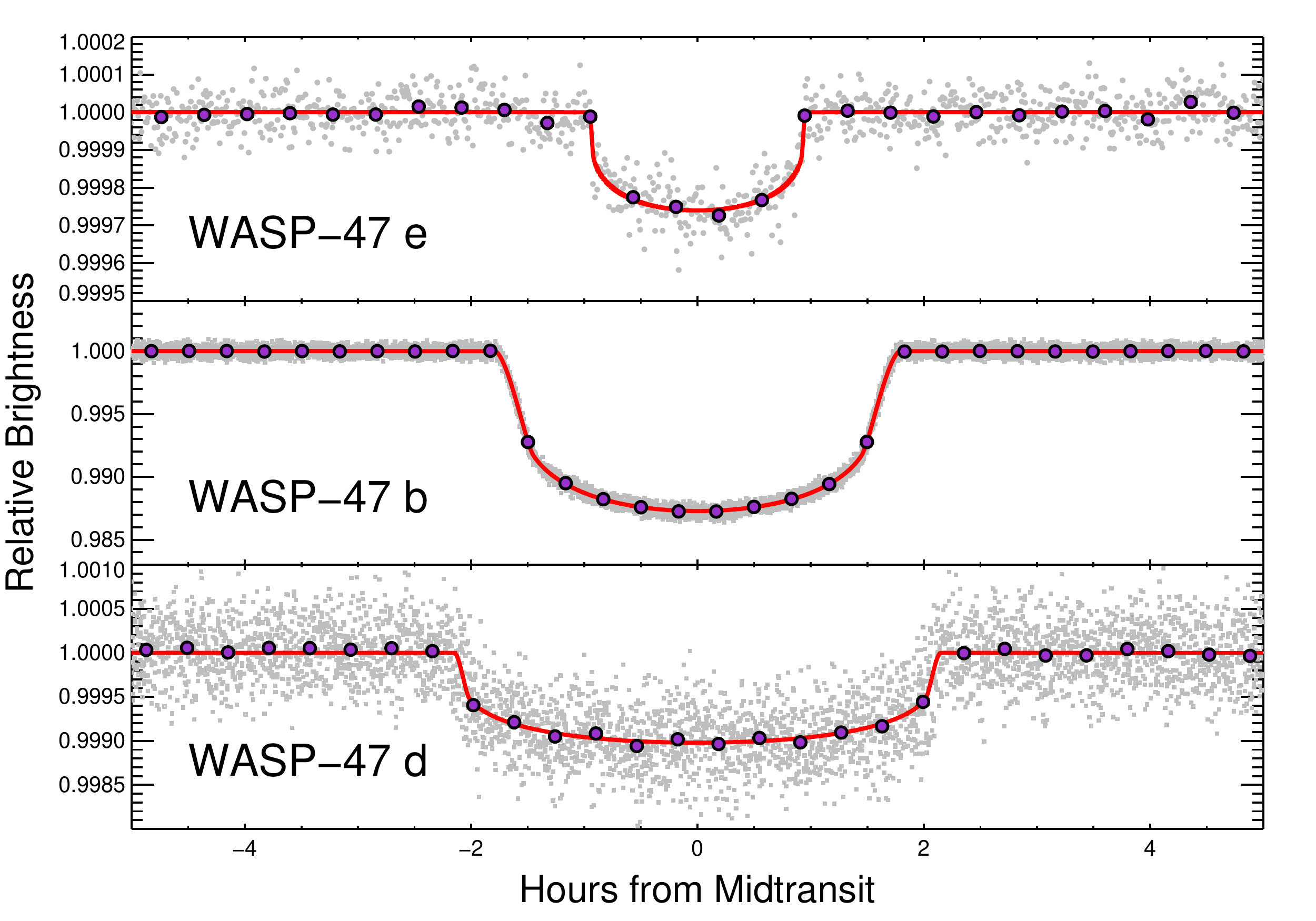}  \hglue-0.34cm
\includegraphics[width=0.44\textwidth]{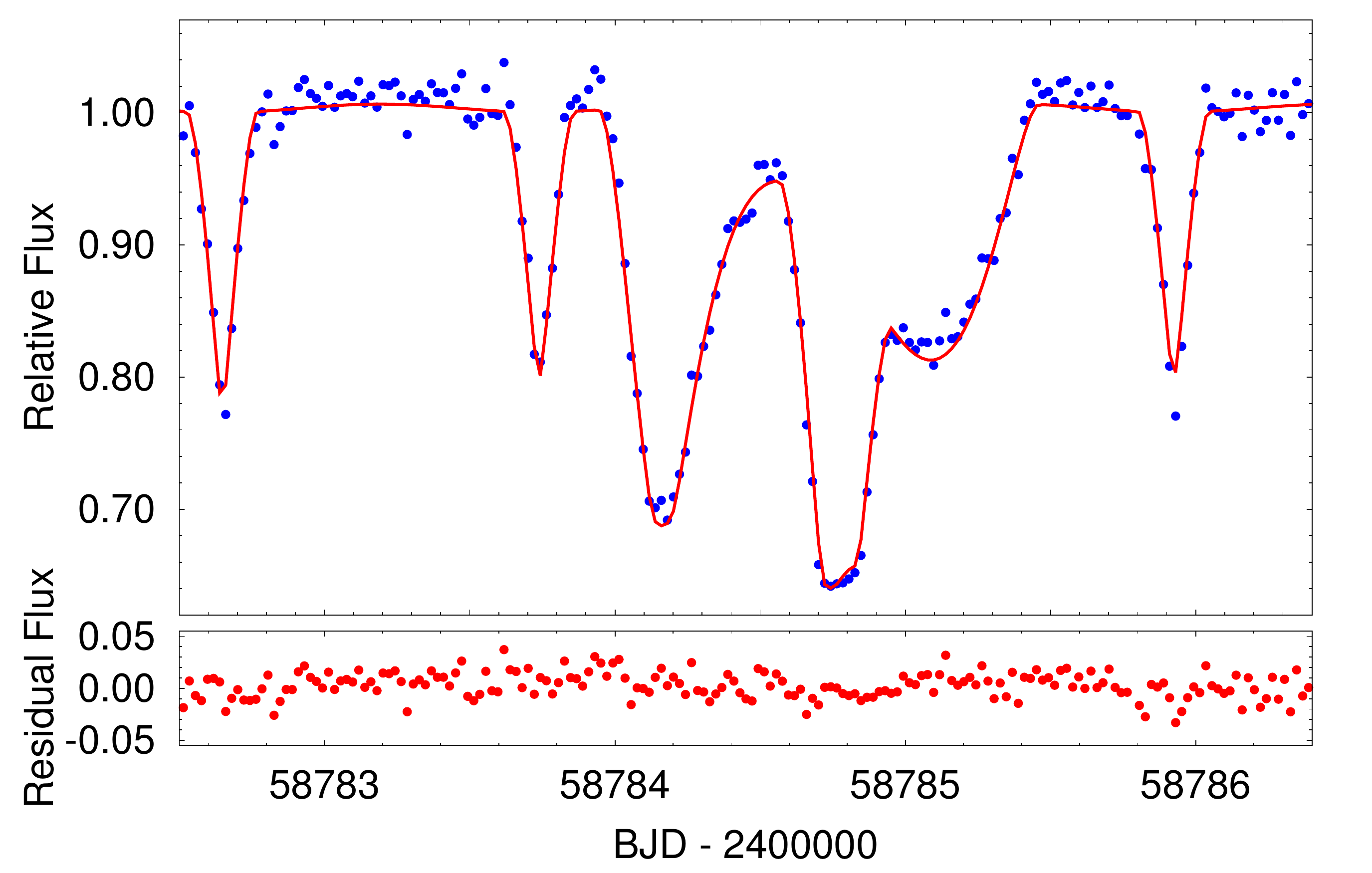} \hglue0.2cm
\includegraphics[width=0.43\textwidth]{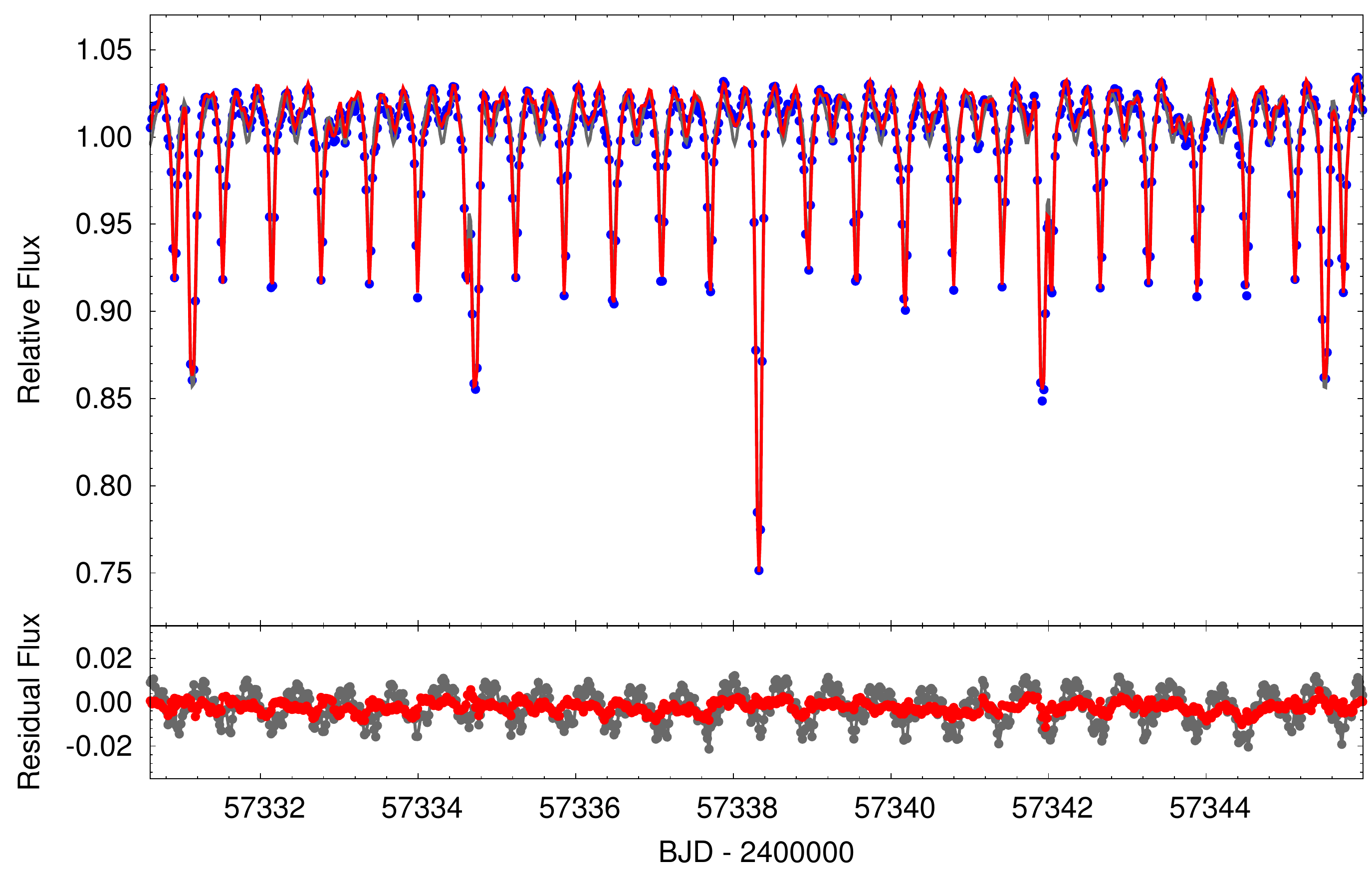} 
\caption{Snapshots of some VSG highlight results. Starting from the first row of panels, and working downward, these are: (1) a mono-transit of the longest period planet found with K2 (EPIC 248847494 b, \cite{giles2018transiting});  (2) transit profiles of the inner three planets of WASP-47 \citep{becker2015wasp}; (3) an outer third-body eclipse of TIC 388459317 \citep{borkovits2022triply}; (4) a quadruple system, EPIC 219217635 \citep{borkovits2018epic}.}
\label{F4}
\end{center}
\end{figure*}

\section{Scientific Impacts} 
The VSG has collectively authored 72 papers (69 peer-reviewed) distributed amongst six main topics: (I) exoplanets, (II) eclipsing binaries (EBs), (III) multistellar systems, including triply eclipsing triple stars, (IV) variable stars, including pulsators and stellar flares, (V) black swans, i.e, unanticipated, new, or rare events, and (VI) dipper stars, i.e, young stars which typically exhibit quasi-regular dipping-flux behavior, which is presumably due to orbiting dusty material. By far, most discoveries made by the VSG are found by several team members within a short time frame. However, there are a few exceptions, where only one VSG-member made the discovery. These cases can be explained by some of us having different personal interests, which thereby creates an intensified lookout for particular types of objects.\\
Usually, the papers are led by one of the four professional astronomers on the team, or a professional astronomer from outside the VSG with whom we regularly collaborate and/or who may be the appropriate expert on the type of source we are reporting. Table~\ref{T2} gives an overview of research topics covered by the VSG including primary and secondary focus topics. The references mentioned from this section represent the VSG's prior and current collaborators in its entirety, and the group's participation in these discoveries is described in what follows.

\begin{table}[h]
\small\sf\centering
\caption{Publication categories for 72 publications authored by the VSG including primary and secondary topics of the papers.\label{T2}}
\begin{tabular}{llll}
\toprule
Topic&Pri.&Sec.&Highlight\\
\midrule
Exoplanets &30 &33 &\cite{vanderburg2016five}\\
Eclipsing binaries &12 &29 &\cite{lacourse2015kepler}\\
Multistellar systems &12 &15 &\cite{powell2021tic}\\
Variable stars &6 &8 &\cite{handler2020tidally}\\
Black swans &7 &9 &\cite{boyajian2016planet}\\
Dipper stars &5 &6 &Capistrant et al. (in-review)\\
\midrule
Total &72 & &Section 4\\
\bottomrule
\end{tabular}\\[10pt]
\end{table}

\begin{figure*}
\begin{center}
\includegraphics[width=0.33 \textwidth, width=0.29 \textheight]{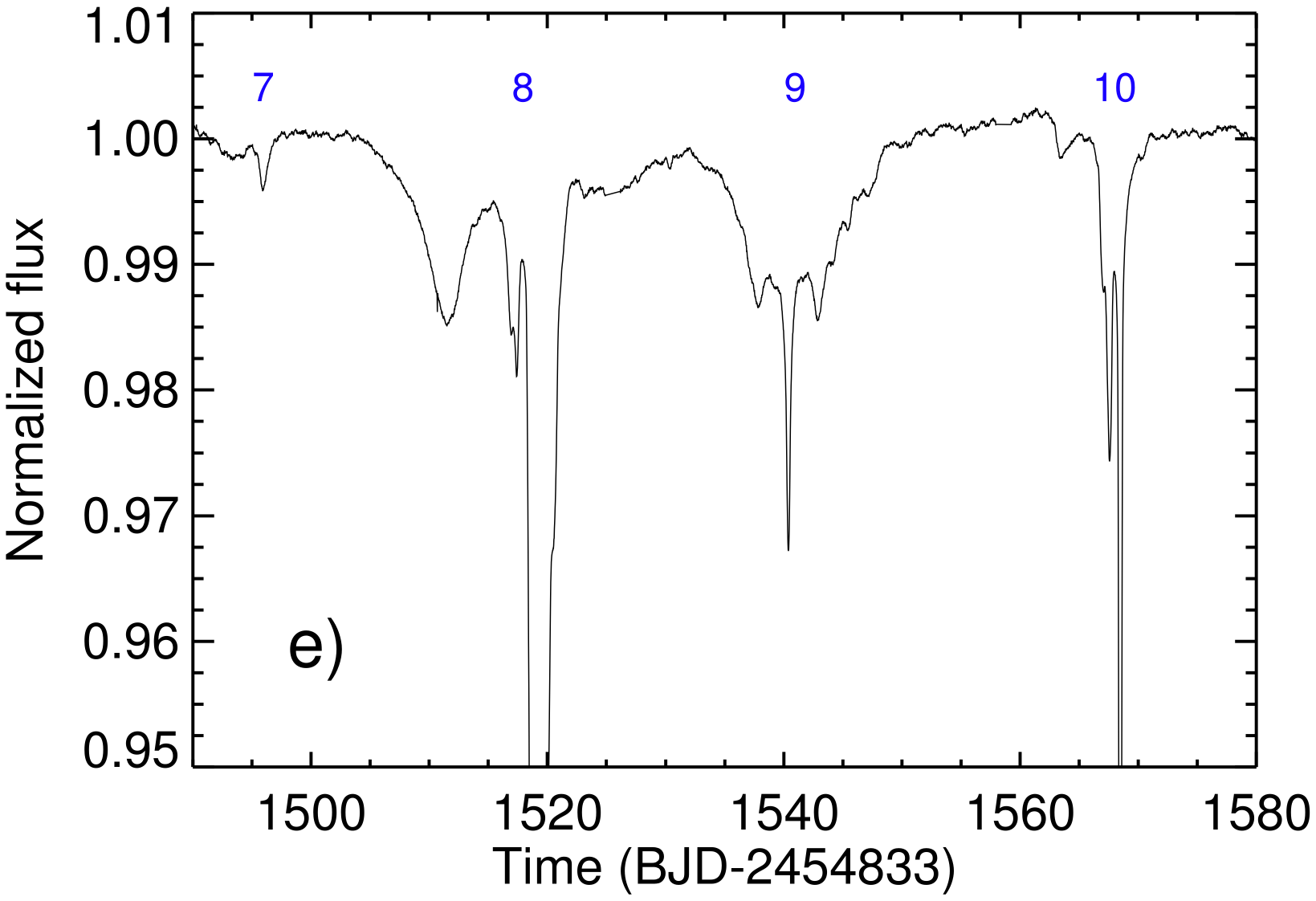} \hglue0.3cm
\includegraphics[width=0.33 \textwidth, width=0.31 \textheight]{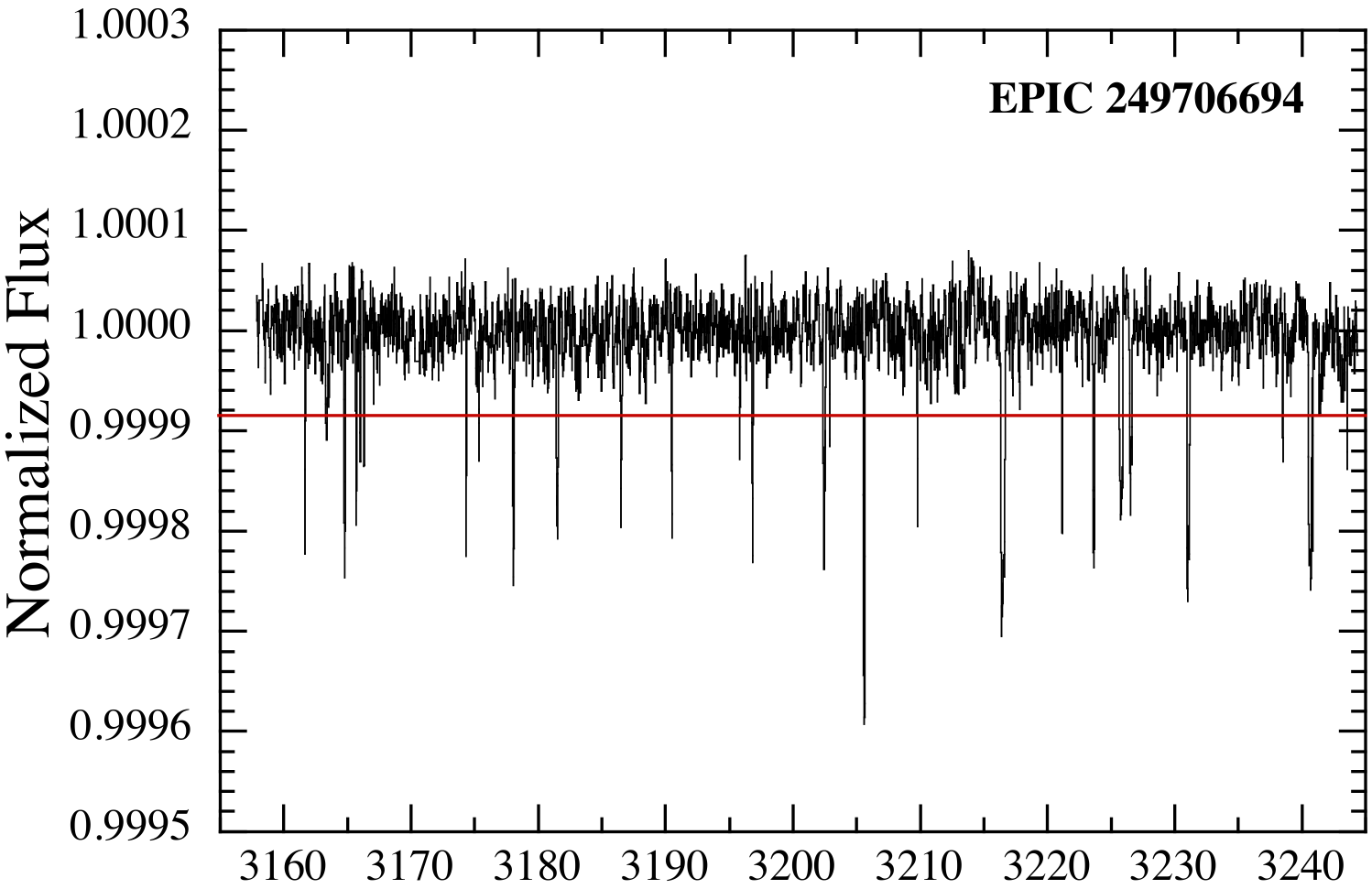} \vglue0.2cm
\includegraphics[width=0.33 \textwidth, width=0.305 \textheight]{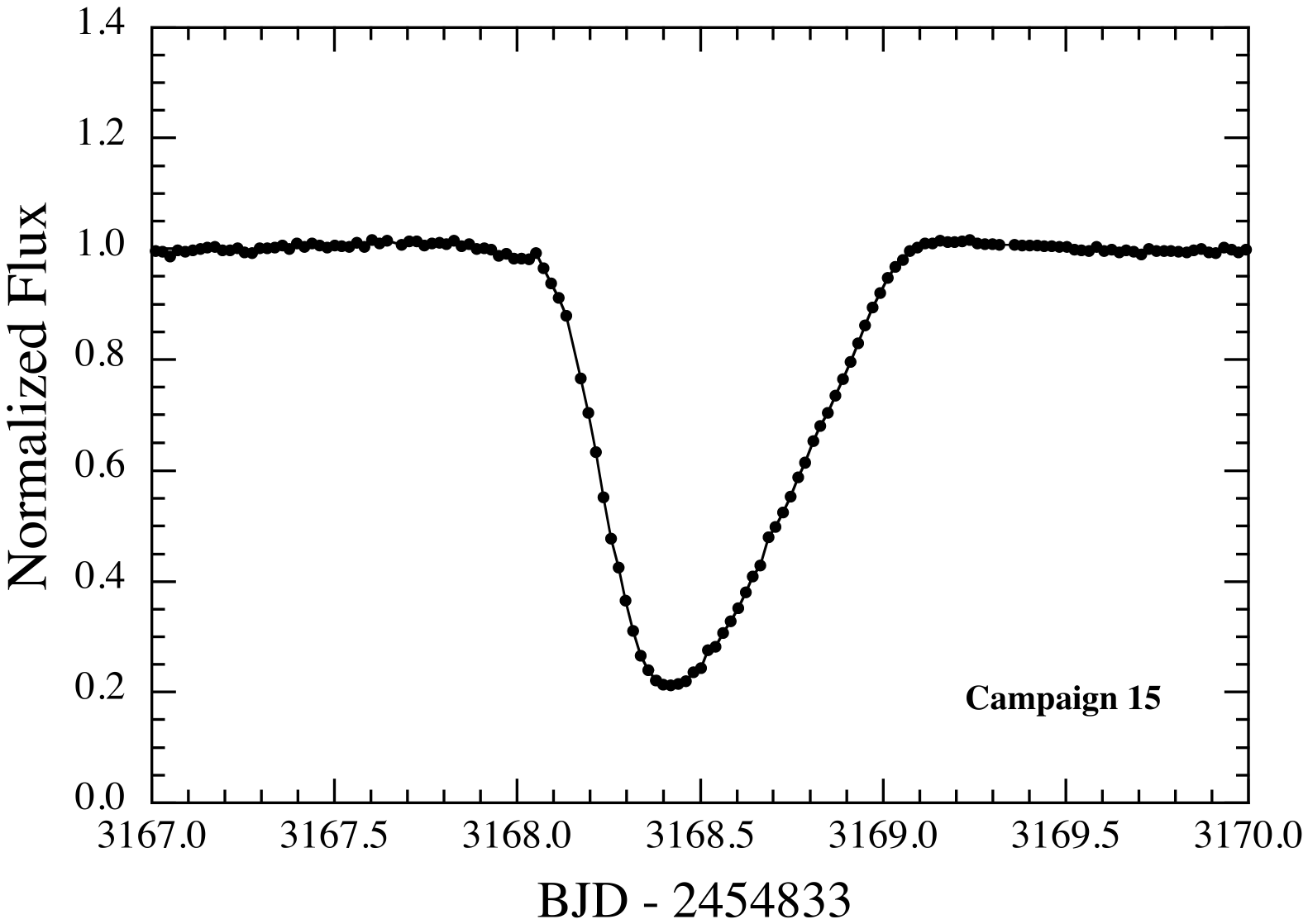} \hglue0.3cm
\includegraphics[width=0.33 \textwidth, width=0.33 
\textheight]{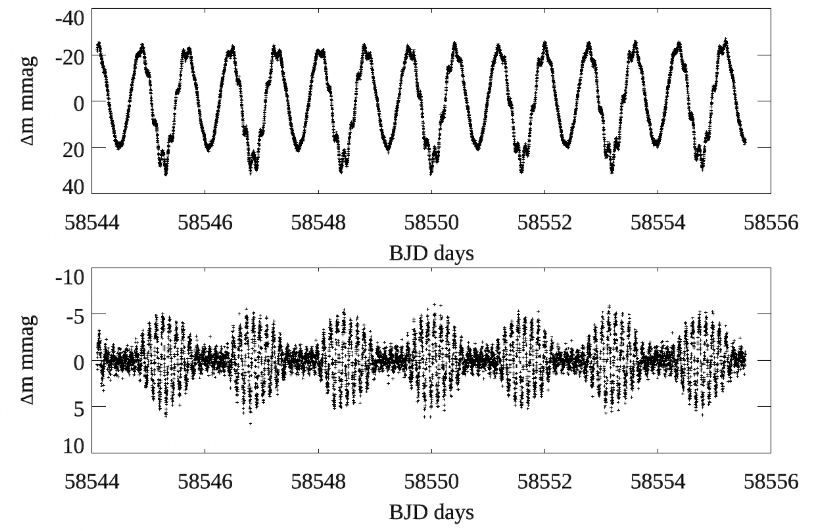} \vglue0.3cm
\includegraphics[width=0.33 \textwidth, width=0.22
\textheight]{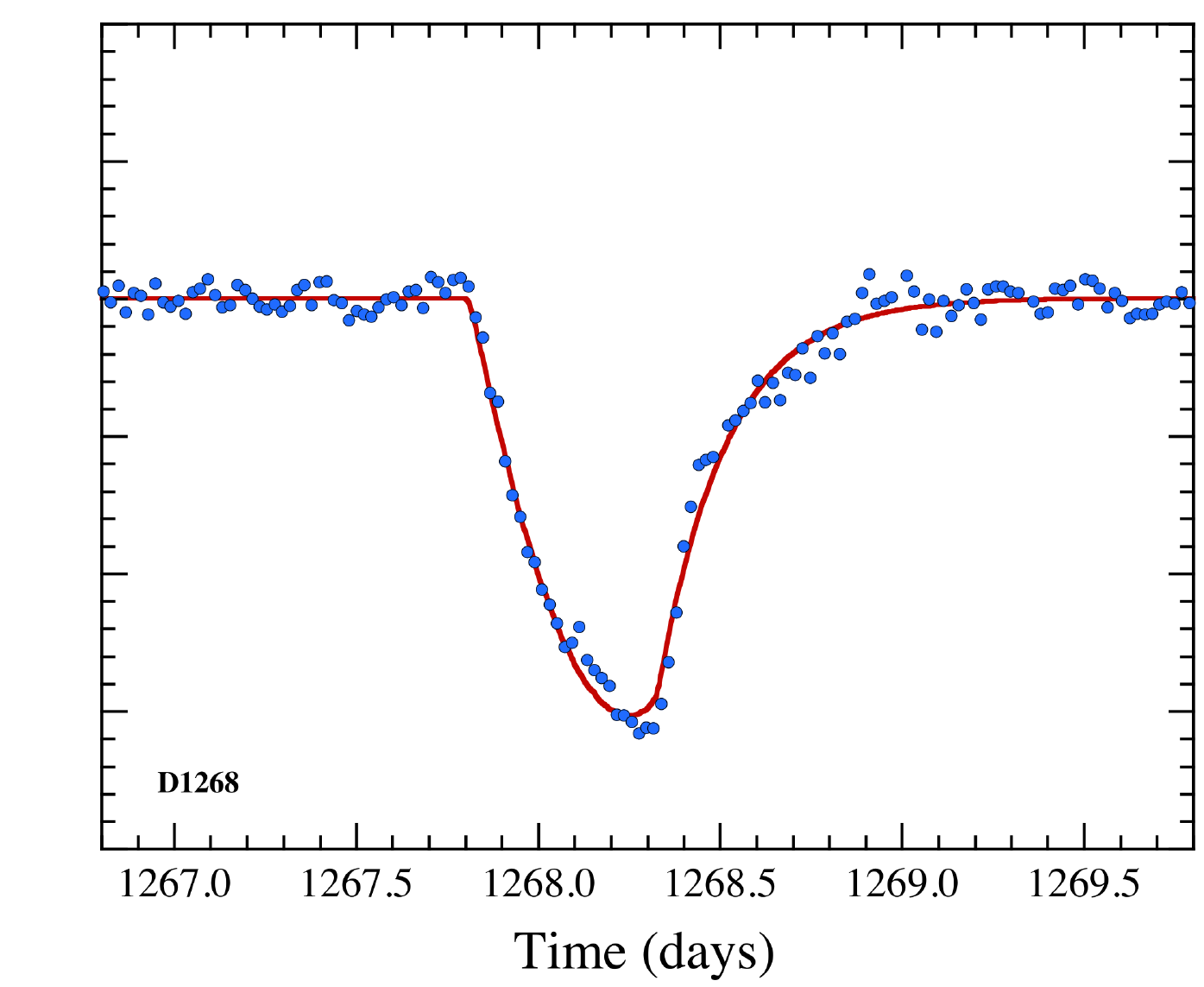} \hglue0.2cm
\includegraphics[width=0.33 \textwidth, width=0.39 \textheight]{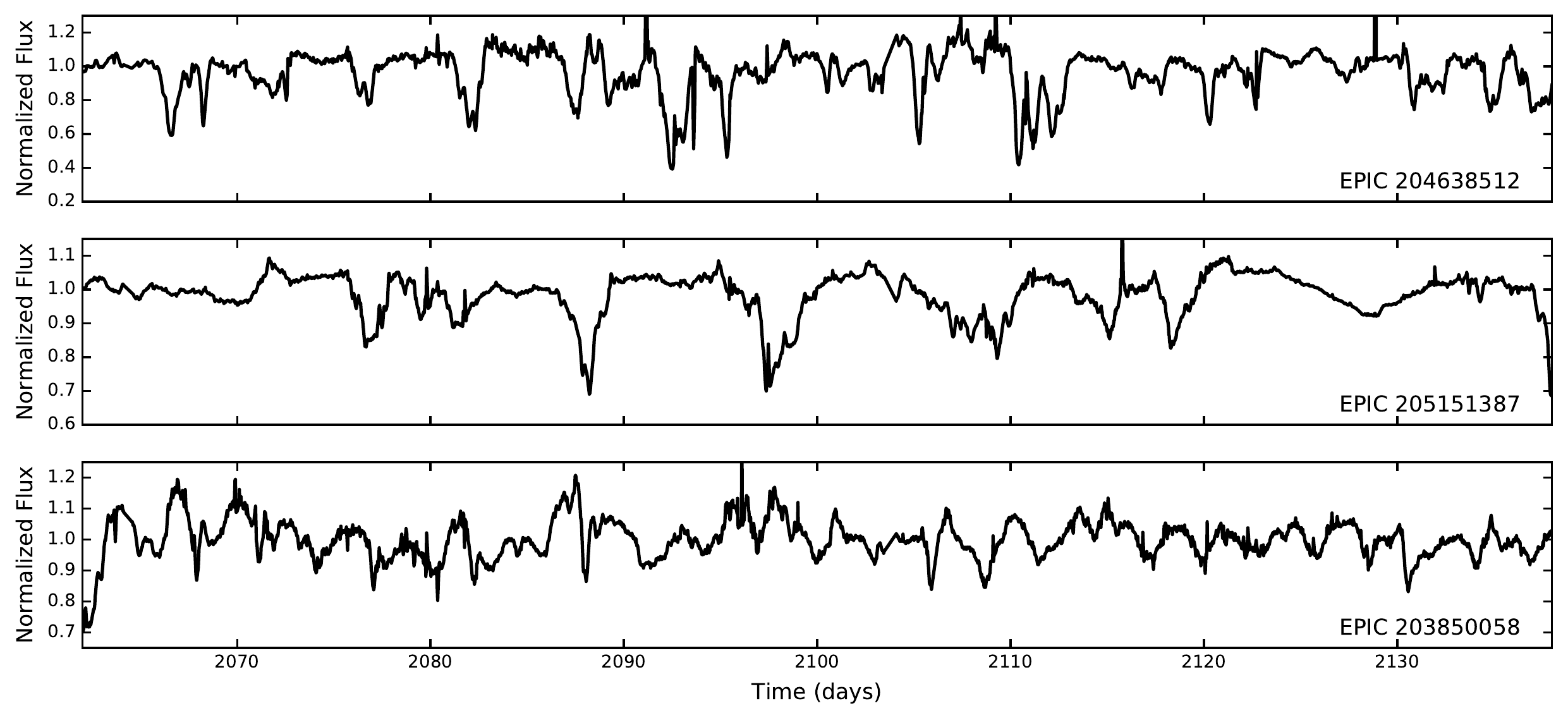} 
\caption{Further snapshots of some VSG highlight results. Starting from the first row of panels, and working downward, these are: (1) a zoom-in on mysterious dips from Boyajian's star \citep{boyajian2016planet}; (2) all 28 dips of the Random Transiter after removal of stellar spot modulations \citep{rappaport2019random}; (3) the asymmetric $\approx$ 80 $\%$ flux occultation (EPIC 204376071, \cite{rappaport2019deep}); (4) stellar pulsator in a binary whose pulsation axis has been tilted to align with the tidal axis of the binary (HD 74423, \cite{handler2020tidally}); (5) a likely exocomet  transit (KIC 3542116, \cite{rappaport2018likely}); (6) light curve profiles of three `dipper' stars (EPICs 204638512, 205151387 and 203850058, \cite{ansdell2016young}).}
\label{F5}
\end{center}
\end{figure*}

\subsection{Exoplanets}
The effort to search for additional, previously undetected exoplanets naturally arose by virtue of the team's involvement in the PH-enterprise. At first, this pursuit was either directed towards single transiting exoplanets (\cite{lintott2013planet}, \cite{wang2013planet}, \cite{wang2013planet2}, \cite{schmitt2015erratum}, \cite{wang2015planet}) and/or multiplanet systems due to the main purpose of the PH-project (\cite{schmitt2014planet2}, \cite{schmitt2014planet}). In addition, contributions were also made in the Exoplanet Explorers project \citep{zink2019catalog}.
The pursuit of lone-signals continued past the close of the collaboration with the original PH science team (\cite{osborn2017long}, \cite{lacourse2018single2}, \cite{quinn2021long}) which led to the discovery of the longest period exoplanet found in K2 (Fig. 4.1 and \cite{giles2018transiting}) and several long-period planets found with TESS photometry (\cite{eisner2020planet}, \cite{dalba2020tess}, \cite{dalba2022tess}, \cite{dalba2022b}).\\
Additionally, the VSG has assisted in discovering several multiple planet systems, including a star hosting three planets \citep{david2018three}, four planets (\cite{daylan2021tess}), five planets (\cite{vanderburg2016five}, \cite{becker2018discrete}, \cite{quinn2019near}) and six planets (\cite{rodriguez2018compact}, \cite{christiansen2018k2}, \cite{hardegree2021k2}, \cite{leleu2021six}).\\
In addition, multistellar systems with orbiting planets were found including a quadruple with a circumbinary planet \citep{schwamb2013planet}, and a second planet in the closest M-dwarf system (triple M-dwarf system) known to host transiting planets \citep{winters2022second}. Also, several lower-order systems at high significance include two planets orbiting a young Sun-like star (\cite{mann2020tess}, \cite{dai2020tess}), a planet around a star in a young star cluster \citep{mann2016zodiacal}, planetary transits for a naked-eye star (V = 5.8) \citep{kane2020transits}, short-period planets \citep{malavolta2018ultra}, and two smaller, inner planets in a known hot jupiter system (Fig. 4.2 and \cite{becker2015wasp}).

\subsection{Eclipsing Binaries}
Although the primary focus by the VSG initially was directed towards exoplanet detections, signals from EBs can resemble that of planets, making a vast collection of EBs a natural byproduct \citep{schmitt2016planet}. This has resulted in large catalogues for \textit{Kepler} \citep{kirk2016kepler}, K2 (\cite{lacourse2015kepler}, \cite{armstrong2015k2}) and \textit{TESS} \citep{prvsa2022tess}. Also, \cite{lacourse2018single2} presented a catalogue of single eclipses found in K2 C0-C14.\\
The VSG also hunts for cataclysmic variable stars (\cite{gies2013kic}, \cite{yu20199}), active, spotted eclipsing giant stars (\cite{olah2018eclipsing}, \cite{olah2020importance}) and EBs with pulsating components (\cite{lee2018eclipsing}, \cite{lee2019pulsating}, \cite{lee2020tess}). Also, a post-Algol system experiencing occultations from an active accretion disk was found \citep{zhou2018occultations}.

\subsection{Multistellar Systems}
In recent years, the VSG has developed a growing interest in hierarchical systems which has resulted in discoveries of triply eclipsing triple systems (\cite{borkovits2019photodynamical}, \cite{borkovits2020compact}, (Fig 4.3 and \cite{borkovits2022triply}), \cite{rappaport2022six}), several quadruple systems where both binaries are eclipsing (Fig. 4.4 and \cite{borkovits2018epic}) including a strongly interacting quadruple \citep{rappaport2017epic}, the nearest known quadruple \citep{borkovits2021bg}, and the most eccentric, low-mass, short-period, eclipsing binary known \citep{han20212+}, which is also part of a quadruple system. Using \textit{TESS}, the VSG presented the largest catalogue of newly discovered doubly eclipsing quadruple candidates \citep{kostov202297}, and also contributed to the PH-TESS discovery of a massive, compact hierarchical system \citep{eisner2022planet}. Adding to this collection, are a quintuple system \citep{rappaport2016quintuple} and the first sextuply eclipsing sextuple star system \citep{powell2021tic}.

\subsection{Variable Stars} 
While searching for exoplanets and EBs, the VSG-members record a variety of others objects, e.g. variable stars \citep{armstrong2015k2}, stars with rigidly rotating magnetospheres \citep{jayaraman2022could} and stellar flares (\cite{gunther2020stellar}, \cite{ilin2021flares}). In this regard, a catalogue of false positive flare signals and newly discovered SSO-candidates is being compiled for a K2-campaign (Kristiansen et al. (in prep.)). During a TESS-survey, the VSG discovered a brand new and important class of pulsators in binary systems called `Tidally Tilted Pulsators' (TTPs). In these systems, the pulsation axis has been tilted into the plane of the binary along the tidal axis. This allows the observer, for the first time, to view a pulsator at aspect angles all the way from 0 to 360 degrees (Fig. 5.4 and \cite{handler2020tidally}, \cite{kurtz2020single}).

\subsection{Black Swans} 
KIC 8462852, also known as Boyajian's star (Fig. 5.1 and \cite{boyajian2016planet}, \cite{boyajian2018first}) was the first black swan discovery with VSG-contribution. The media quickly dubbed it an alien megastructure by way of inspiration from \cite{wright2015g}. Despite the apparent rarity of the system (a properly comparable analog has yet to be discovered in Kepler, K2 or TESS data), subsequent ground-based observations have established that dusty material is the most likely explanation for the irregular sharp variations in flux (see e.g. \cite{hitchcock2019non}). Also found with Kepler photometry, \cite{rappaport2018likely} reported the first transit signals likely caused by extra solar comets (Fig. 5.5). This represented a major step forward in terms of detecting very small objects orbiting stars via a trail of dust emissions. With K2, an interesting M-dwarf with a deep, asymmetric drop in flux $\approx$ 80 $\%$ was discovered (Fig. 5.3 and \cite{rappaport2019deep}), and a Sun-like star experiencing 28 transit-like dips of similar depths but showing no periodicity (Fig. 5.2 and \cite{rappaport2019random}). More recently, \textit{TESS} has revealed a mysterious dust-emitting object that orbits its host star every 20 days \citep{powell2021mysterious}, complex and rapidly rotating M-dwarfs \citep{gunther2020complex}, and an unanticipated find of TTPs (Sect. 4.4, Fig. 5.4).

\subsection{Dipper Stars}
Early in the repurposed K2-mission, when \textit{Kepler} was pointed towards the ecliptic plane, the first K2-dippers surfaced (Fig. 5.6 and \cite{ansdell2016young}, \cite{ansdell2016dipper}). These stars are young with dusty orbiting material showing irregular flux variations in their light curve profiles. These stars are of particular interest in the regime of planet formation scenarios \citep{gaidos2019planetesimals}. Although dipper stars' flux variations frequently are deep, shallow dipper-events have also been recorded \citep{ansdell2019little}. Using \textit{TESS} photometry, Capistrant et al. (in-review) presents the largest catalogue of dippers to date and thereby double the known dipper population. Most of the dippers in this catalogue were found by the VSG.

\section{Discussion and Conclusion} 
In this work, we have presented the Visual Survey Group (VSG) and its Pro-Am collaborative nature, including its history, survey-methods and discoveries. Over the past decade, the VSG has collectively surveyed nearly 10 million light curves manually and authored 69 peer-reviewed papers primarily focusing on exoplanets, eclipsing multistellar objects and `black swans'. However, the quantity of data produced by \textit{TESS} is too immense to keep up a completely updated, thorough and manual search. Although the manual search-method by far has been the dominant approach used by the VSG, a combination of automated and manual searches is becoming more frequent. This said, no group members intend to phase out the manual search since it not only complements the prevailing approaches but simultaneously is able to reveal some of the hidden gems of our local Universe.\\
Concerning manual searches, there are several limitations worth mentioning.  These include eye fatigue, light curve scrolling misses due to classification speed, light curve size, i.e number of data points, light curve presentation degradation caused by computer monitor resolution, and simply the loss of focus which most frequently occurs after numerous hours of surveying. By way of example, no additional stars similar to Boyajian's star are expected to hide in the Kepler-data set, but additional shallow exocomets may have been missed, because they are isolated and hard to distinguish in a full light curve. Likewise, manual surveys do not perform particularly well for small exoplanets buried in the noise floor.  Also, the human eye is not very good at seeing very short period (comparable to the sampling time) periodic events with low signal to noise, which can be easy to find with various periodogram searches (e.g., Fourier transforms, BLS transforms). \\
On the other hand, automated programs are not particularly good at recognizing new object patterns, and might thereby discard new and interesting types of phenomena. Several of our findings have illustrated this effect.  Overall, we have found that manual surveys nicely complement automated searches and are able to complete occurrence rate studies. In addition, years of experience with manual surveying makes light curve artefacts stand out more and thereby reduces wasted time and effort in tracking down spurious signals that relatively new surveyors might not recognize. \\
Finally, we invite other researchers to contact us with the purpose of collaboration in mind. We are very open to looking for particular classes of objects that we might not otherwise have paid attention to, but which others find quite interesting.  Also, we welcome other experienced surveyors to join the hunt with us in the VSG.

\begin{acks}
We thank the anonymous referee for feedback which has significantly improved the clarity of the paper. We very much appreciate help and advice from, and consultations with, various professional astronomers in different sub-fields, including Tamas Borkovits, Hugh Osborn, Eric Gaidos, Jonathan Labadie-Bartz, Don Kurtz, Gerald Handler, Rahul Jayaraman, Chelsea Huang, Maximilian G\"unther, Melinda Soares-Furtado, Joey Rodriguez, Paul Dalba, Katalin Ol\'ah, Jae Woo Lee, Andrej Pr\v{s}a, Petr Zasche, Pierre Maxted, Andrei Tokovinin, Ekaterina Ilin, Jennifer Winters, Andrew Mann, Megan Ansdell, Kevin Hardegree-Ullman, Adrien Leleu, Samuel Quinn, Tansu Daylan, Luca Malavolta, David Armstrong and the former PH-science team. This paper includes data collected by the Kepler mission and obtained from the MAST data archive at the Space Telescope Science Institute (STScI). Funding for the Kepler mission is provided by the NASA Science Mission Directorate. STScI is operated by the Association of Universities for Research in Astronomy, Inc., under NASA contract NAS 5–26555. This paper includes data collected with the TESS mission, obtained from the MAST data archive at the Space Telescope Science Institute (STScI). Funding for the TESS mission is provided by the NASA Explorer Program. STScI is operated by the Association of Universities for Research in Astronomy, Inc., under NASA contract NAS 5–26555. This research has made use of the Exoplanet Follow-up Observation Program website, which is operated by the California Institute of Technology, under contract with the National Aeronautics and Space Administration under the Exoplanet Exploration Program. This research made use of Lightkurve, a Python package for Kepler and TESS data analysis (Lightkurve Collaboration, 2018). This research has made use of the SIMBAD database, operated at CDS, Strasbourg, France. This work has made use of data from the European Space Agency (ESA) mission
\textit{Gaia} (\url{https://www.cosmos.esa.int/gaia}), processed by the \textit{Gaia}
Data Processing and Analysis Consortium (DPAC,
\url{https://www.cosmos.esa.int/web/gaia/dpac/consortium}). Funding for the DPAC
has been provided by national institutions, in particular the institutions
participating in the \textit{Gaia} Multilateral Agreement. This research has made use data obtained by WISE of the NASA/IPAC Infrared Science Archive, which is funded by the National Aeronautics and Space Administration and operated by the California Institute of Technology.
\end{acks}

\begin{funding}
This research received no specific grant from any funding agency in the public, commercial, or not-for-profit sectors.
\end{funding}

\theendnotes

\bibliographystyle{SageH}
\bibliography{vsg.bbl}

\begin{thebibliography}{137}
\providecommand{\natexlab}[1]{#1}
\providecommand{\url}[1]{\texttt{#1}}
\providecommand{\urlprefix}{URL }
\expandafter\ifx\csname urlstyle\endcsname\relax
  \providecommand{\doi}[1]{DOI:\discretionary{}{}{}#1}\else
  \providecommand{\doi}{DOI:\discretionary{}{}{}\begingroup
  \urlstyle{rm}\Url}\fi

\bibitem[{Agol and Fabrycky(2018)}]{agol2018transit}
Agol E and Fabrycky D (2018) Transit-timing and duration variations for the
  discovery and characterization of exoplanets, 7, doi: 10.1007 .

\bibitem[{Anderson et~al.(2000)Anderson, Werthimer, Cobb, Korpela, Lebofsky,
  Gedye and Sullivan}]{anderson2000seti}
Anderson D, Werthimer D, Cobb J, Korpela E, Lebofsky M, Gedye D and Sullivan WT
  (2000) Seti@ home: internet distributed computing for seti.
\newblock In: \emph{Bioastronomy 99}, volume 213.

\bibitem[{Ansdell et~al.(2019)Ansdell, Gaidos, Jacobs, Mann, Manara, Kennedy,
  Vanderburg, Kenworthy, Hirano, LaCourse et~al.}]{ansdell2019little}
Ansdell M, Gaidos E, Jacobs TL, Mann A, Manara CF, Kennedy GM, Vanderburg A,
  Kenworthy M, Hirano T, LaCourse DM et~al. (2019) The little dippers: transits
  of star-grazing exocomets?
\newblock \emph{Monthly Notices of the Royal Astronomical Society} 483(3):
  3579--3591.

\bibitem[{Ansdell et~al.(2016{\natexlab{a}})Ansdell, Gaidos, Rappaport, Jacobs,
  LaCourse, Jek, Mann, Wyatt, Kennedy, Williams et~al.}]{ansdell2016young}
Ansdell M, Gaidos E, Rappaport S, Jacobs T, LaCourse D, Jek K, Mann A, Wyatt M,
  Kennedy G, Williams J et~al. (2016{\natexlab{a}}) Young “dipper” stars in
  upper sco and oph observed by k2.
\newblock \emph{The Astrophysical Journal} 816(2): 69.

\bibitem[{Ansdell et~al.(2016{\natexlab{b}})Ansdell, Gaidos, Williams, Kennedy,
  Wyatt, LaCourse, Jacobs and Mann}]{ansdell2016dipper}
Ansdell M, Gaidos E, Williams J, Kennedy G, Wyatt M, LaCourse D, Jacobs T and
  Mann A (2016{\natexlab{b}}) Dipper discs not inclined towards edge-on orbits.
\newblock \emph{Monthly Notices of the Royal Astronomical Society: Letters}
  462(1): L101--L105.

\bibitem[{Armstrong et~al.(2015)Armstrong, Kirk, Lam, McCormac, Osborn, Spake,
  Walker, Brown, Kristiansen, Pollacco et~al.}]{armstrong2015k2}
Armstrong D, Kirk J, Lam K, McCormac J, Osborn H, Spake J, Walker S, Brown D,
  Kristiansen M, Pollacco D et~al. (2015) K2 variable catalogue--ii. machine
  learning classification of variable stars and eclipsing binaries in k2 fields
  0--4.
\newblock \emph{Monthly Notices of the Royal Astronomical Society} 456(2):
  2260--2272.

\bibitem[{Barentsen and Cardoso(2018)}]{barentsen2018kadenza}
Barentsen G and Cardoso JVdM (2018) Kadenza: Kepler/k2 raw cadence data reader.
\newblock \emph{Astrophysics Source Code Library} : ascl--1803.

\bibitem[{Batalha et~al.(2013)Batalha, Rowe, Bryson, Barclay, Burke, Caldwell,
  Christiansen, Mullally, Thompson, Brown et~al.}]{batalha2013planetary}
Batalha NM, Rowe JF, Bryson ST, Barclay T, Burke CJ, Caldwell DA, Christiansen
  JL, Mullally F, Thompson SE, Brown TM et~al. (2013) Planetary candidates
  observed by kepler. iii. analysis of the first 16 months of data.
\newblock \emph{The Astrophysical Journal Supplement Series} 204(2): 24.

\bibitem[{Becker et~al.(2015)Becker, Vanderburg, Adams, Rappaport and
  Schwengeler}]{becker2015wasp}
Becker JC, Vanderburg A, Adams FC, Rappaport SA and Schwengeler HM (2015)
  Wasp-47: A hot jupiter system with two additional planets discovered by k2.
\newblock \emph{The Astrophysical Journal Letters} 812(2): L18.

\bibitem[{Becker et~al.(2018)Becker, Vanderburg, Rodriguez, Omohundro, Adams,
  Stassun, Yao, Hartman, Pepper, Bakos et~al.}]{becker2018discrete}
Becker JC, Vanderburg A, Rodriguez JE, Omohundro M, Adams FC, Stassun KG, Yao
  X, Hartman J, Pepper J, Bakos G et~al. (2018) A discrete set of possible
  transit ephemerides for two long-period gas giants orbiting hip 41378.
\newblock \emph{The Astronomical Journal} 157(1): 19.

\bibitem[{Berthier et~al.(2016)Berthier, Carry, Vachier, Eggl and
  Santerne}]{berthier2016prediction}
Berthier J, Carry B, Vachier F, Eggl S and Santerne A (2016) Prediction of
  transits of solar system objects in kepler/k2 images: An extension of the
  virtual observatory service skybot.
\newblock \emph{Monthly Notices of the Royal Astronomical Society} 458(3):
  3394--3398.

\bibitem[{Berthier et~al.(2006)Berthier, Vachier, Thuillot, Fernique,
  Ochsenbein, Genova, Lainey and Arlot}]{berthier2006skybot}
Berthier J, Vachier F, Thuillot W, Fernique P, Ochsenbein F, Genova F, Lainey V
  and Arlot JE (2006) Skybot, a new vo service to identify solar system
  objects.
\newblock In: \emph{Astronomical Data Analysis Software and Systems XV}, volume
  351. p. 367.

\bibitem[{Borkovits et~al.(2018)Borkovits, Albrecht, Rappaport, Nelson,
  Vanderburg, Gary, Tan, Justesen, Kristiansen, Jacobs
  et~al.}]{borkovits2018epic}
Borkovits T, Albrecht S, Rappaport S, Nelson L, Vanderburg A, Gary B, Tan T,
  Justesen A, Kristiansen M, Jacobs T et~al. (2018) Epic 219217635: a doubly
  eclipsing quadruple system containing an evolved binary.
\newblock \emph{Monthly Notices of the Royal Astronomical Society} 478(4):
  5135--5152.

\bibitem[{Borkovits et~al.(2022)Borkovits, Mitnyan, Rappaport, Pribulla,
  Powell, Kostov, B{\'\i}r{\'o}, Cs{\'a}nyi, Garai, Gary
  et~al.}]{borkovits2022triply}
Borkovits T, Mitnyan T, Rappaport S, Pribulla T, Powell B, Kostov V,
  B{\'\i}r{\'o} I, Cs{\'a}nyi I, Garai Z, Gary B et~al. (2022) Triply eclipsing
  triple stars in the northern tess fields: Tics 193993801, 388459317, and
  52041148.
\newblock \emph{Monthly Notices of the Royal Astronomical Society} 510(1):
  1352--1374.

\bibitem[{Borkovits et~al.(2019)Borkovits, Rappaport, Kaye, Isaacson,
  Vanderburg, Howard, Kristiansen, Omohundro, Schwengeler, Terentev
  et~al.}]{borkovits2019photodynamical}
Borkovits T, Rappaport S, Kaye T, Isaacson H, Vanderburg A, Howard A,
  Kristiansen M, Omohundro M, Schwengeler H, Terentev I et~al. (2019)
  Photodynamical analysis of the triply eclipsing hierarchical triple system
  epic 249432662.
\newblock \emph{Monthly Notices of the Royal Astronomical Society} 483(2):
  1934--1951.

\bibitem[{Borkovits et~al.(2021)Borkovits, Rappaport, Maxted, Terentev,
  Omohundro, Gagliano, Jacobs, Kristiansen, LaCourse, Schwengeler
  et~al.}]{borkovits2021bg}
Borkovits T, Rappaport S, Maxted P, Terentev I, Omohundro M, Gagliano R, Jacobs
  T, Kristiansen M, LaCourse D, Schwengeler H et~al. (2021) Bg ind: the nearest
  doubly eclipsing, compact hierarchical quadruple system.
\newblock \emph{Monthly Notices of the Royal Astronomical Society} 503(3):
  3759--3774.

\bibitem[{Borkovits et~al.(2020)Borkovits, Rappaport, Tan, Gagliano, Jacobs,
  Huang, Mitnyan, Hambsch, Kaye, Maxted et~al.}]{borkovits2020compact}
Borkovits T, Rappaport S, Tan T, Gagliano R, Jacobs T, Huang X, Mitnyan T,
  Hambsch F, Kaye T, Maxted P et~al. (2020) The compact triply eclipsing triple
  star tic 209409435 discovered with tess.
\newblock \emph{Monthly Notices of the Royal Astronomical Society} 496(4):
  4624--4636.

\bibitem[{Borucki et~al.(2010)Borucki, Koch, Basri, Batalha, Brown, Caldwell,
  Caldwell, Christensen-Dalsgaard, Cochran, DeVore et~al.}]{borucki2010kepler}
Borucki WJ, Koch D, Basri G, Batalha N, Brown T, Caldwell D, Caldwell J,
  Christensen-Dalsgaard J, Cochran WD, DeVore E et~al. (2010) Kepler
  planet-detection mission: introduction and first results.
\newblock \emph{Science} 327(5968): 977--980.

\bibitem[{Bouma et~al.(2019)Bouma, Hartman, Bhatti, Winn and
  Bakos}]{bouma2019cluster}
Bouma L, Hartman J, Bhatti W, Winn J and Bakos G (2019) Cluster difference
  imaging photometric survey. i. light curves of stars in open clusters from
  tess sectors 6 and 7.
\newblock \emph{The Astrophysical Journal Supplement Series} 245(1): 13.

\bibitem[{Boyajian et~al.(2018)Boyajian, Alonso, Ammerman, Armstrong, Ramos,
  Barkaoui, Beatty, Benkhaldoun, Benni, Bentley et~al.}]{boyajian2018first}
Boyajian TS, Alonso R, Ammerman A, Armstrong D, Ramos AA, Barkaoui K, Beatty
  TG, Benkhaldoun Z, Benni P, Bentley RO et~al. (2018) The first post-kepler
  brightness dips of kic 8462852.
\newblock \emph{The Astrophysical Journal Letters} 853(1): L8.

\bibitem[{Boyajian et~al.(2016)Boyajian, LaCourse, Rappaport, Fabrycky,
  Fischer, Gandolfi, Kennedy, Korhonen, Liu, Moor et~al.}]{boyajian2016planet}
Boyajian TS, LaCourse D, Rappaport S, Fabrycky D, Fischer D, Gandolfi D,
  Kennedy G, Korhonen H, Liu M, Moor A et~al. (2016) Planet hunters ix. kic
  8462852--where's the flux?
\newblock \emph{Monthly Notices of the Royal Astronomical Society} 457(4):
  3988--4004.

\bibitem[{Boyd(2011)}]{boyd2011pro}
Boyd D (2011) Pro-am collaboration in astronomy-past, present and future.
\newblock \emph{Journal of the British Astronomical Association} 121: 73--90.

\bibitem[{Brasseur et~al.(2019)Brasseur, Phillip, Fleming, Mullally and
  White}]{brasseur2019astrocut}
Brasseur C, Phillip C, Fleming SW, Mullally S and White RL (2019) Astrocut:
  Tools for creating cutouts of tess images.
\newblock \emph{Astrophysics Source Code Library} : ascl--1905.

\bibitem[{Brown et~al.(2021)Brown, Vallenari, Prusti, De~Bruijne, Babusiaux,
  Biermann, Creevey, Evans, Eyer, Hutton et~al.}]{brown2021gaia}
Brown AG, Vallenari A, Prusti T, De~Bruijne J, Babusiaux C, Biermann M, Creevey
  O, Evans D, Eyer L, Hutton A et~al. (2021) Gaia early data release 3-summary
  of the contents and survey properties.
\newblock \emph{Astronomy \& Astrophysics} 649: A1.

\bibitem[{Brown et~al.(2011)Brown, Latham, Everett and
  Esquerdo}]{brown2011kepler}
Brown TM, Latham DW, Everett ME and Esquerdo GA (2011) Kepler input catalog:
  Photometric calibration and stellar classification.
\newblock \emph{The Astronomical Journal} 142(4): 112.

\bibitem[{Caldwell et~al.(2020)Caldwell, Tenenbaum, Twicken, Jenkins, Ting,
  Smith, Hedges, Fausnaugh, Rose and Burke}]{caldwell2020tess}
Caldwell DA, Tenenbaum P, Twicken JD, Jenkins JM, Ting E, Smith JC, Hedges C,
  Fausnaugh MM, Rose M and Burke C (2020) Tess science processing operations
  center ffi target list products.
\newblock \emph{Research Notes of the AAS} 4(11): 201.

\bibitem[{Chambers et~al.(2016)Chambers, Magnier, Metcalfe, Flewelling, Huber,
  Waters, Denneau, Draper, Farrow, Finkbeiner et~al.}]{chambers2016pan}
Chambers KC, Magnier E, Metcalfe N, Flewelling H, Huber M, Waters C, Denneau L,
  Draper P, Farrow D, Finkbeiner D et~al. (2016) The pan-starrs1 surveys.
\newblock \emph{arXiv preprint arXiv:1612.05560} .

\bibitem[{Christiansen et~al.(2018)Christiansen, Crossfield, Barentsen,
  Lintott, Barclay, Simmons, Petigura, Schlieder, Dressing, Vanderburg
  et~al.}]{christiansen2018k2}
Christiansen JL, Crossfield IJ, Barentsen G, Lintott CJ, Barclay T, Simmons BD,
  Petigura E, Schlieder JE, Dressing CD, Vanderburg A et~al. (2018) The k2-138
  system: A near-resonant chain of five sub-neptune planets discovered by
  citizen scientists.
\newblock \emph{The Astronomical Journal} 155(2): 57.

\bibitem[{Collaboration and Wilkinson(2016)}]{collaboration2016gaia}
Collaboration G and Wilkinson M (2016) Gaia data release 1: Summary of the
  astrometric, photometric, and survey properties .

\bibitem[{Cooke et~al.(2018)Cooke, Pollacco, West, McCormac and
  Wheatley}]{cooke2018single}
Cooke BF, Pollacco D, West R, McCormac J and Wheatley PJ (2018) Single site
  observations of tess single transit detections.
\newblock \emph{Astronomy \& Astrophysics} 619: A175.

\bibitem[{Cui et~al.(2021)Cui, Liu, Feng and Liu}]{cui2021identify}
Cui K, Liu J, Feng F and Liu J (2021) Identify light-curve signals with deep
  learning based object detection algorithm. i. transit detection.
\newblock \emph{The Astronomical Journal} 163(1): 23.

\bibitem[{Dai et~al.(2020)Dai, Roy, Fulton, Robertson, Hirsch, Isaacson,
  Albrecht, Mann, Kristiansen, Batalha et~al.}]{dai2020tess}
Dai F, Roy A, Fulton B, Robertson P, Hirsch L, Isaacson H, Albrecht S, Mann AW,
  Kristiansen MH, Batalha NM et~al. (2020) The tess-keck survey. iii. a stellar
  obliquity measurement of toi-1726 c.
\newblock \emph{The Astronomical Journal} 160(4): 193.

\bibitem[{Dalba et~al.(2020)Dalba, Gupta, Rodriguez, Dragomir, Huang, Kane,
  Quinn, Bieryla, Esquerdo, Fulton et~al.}]{dalba2020tess}
Dalba PA, Gupta AF, Rodriguez JE, Dragomir D, Huang CX, Kane SR, Quinn SN,
  Bieryla A, Esquerdo GA, Fulton BJ et~al. (2020) The tess--keck survey. i. a
  warm sub-saturn-mass planet and a caution about stray light in tess cameras.
\newblock \emph{The Astronomical Journal} 159(5): 241.

\bibitem[{Dalba et~al.(2022{\natexlab{a}})Dalba, Jacobs, Omohundro, Gagliano,
  Jursich, Kristiansen, LaCourse, Schwengeler and Terentev}]{dalba2022b}
Dalba PA, Jacobs TL, Omohundro M, Gagliano R, Jursich J, Kristiansen MH,
  LaCourse DM, Schwengeler HM and Terentev IA (2022{\natexlab{a}}) The refined
  transit ephemeris of toi-2180 b.
\newblock \emph{Research Notes of the AAS} 6(4).

\bibitem[{Dalba et~al.(2022{\natexlab{b}})Dalba, Kane, Dragomir, Villanueva,
  Collins, Jacobs, LaCourse, Gagliano, Kristiansen, Omohundro
  et~al.}]{dalba2022tess}
Dalba PA, Kane SR, Dragomir D, Villanueva S, Collins KA, Jacobs TL, LaCourse
  DM, Gagliano R, Kristiansen MH, Omohundro M et~al. (2022{\natexlab{b}}) The
  tess-keck survey. viii. confirmation of a transiting giant planet on an
  eccentric 261 day orbit with the automated planet finder telescope.
\newblock \emph{The Astronomical Journal} 163(2): 61.

\bibitem[{David et~al.(2018)David, Crossfield, Benneke, Petigura, Gonzales,
  Schlieder, Yu, Isaacson, Howard, Ciardi et~al.}]{david2018three}
David TJ, Crossfield IJ, Benneke B, Petigura EA, Gonzales EJ, Schlieder JE, Yu
  L, Isaacson HT, Howard AW, Ciardi DR et~al. (2018) Three small planets
  transiting the bright young field star k2-233.
\newblock \emph{The Astronomical Journal} 155(5): 222.

\bibitem[{Daylan et~al.(2021)Daylan, Pingl{\'e}, Wright, G{\"u}nther, Stassun,
  Kane, Vanderburg, Jontof-Hutter, Rodriguez, Shporer et~al.}]{daylan2021tess}
Daylan T, Pingl{\'e} K, Wright J, G{\"u}nther MN, Stassun KG, Kane SR,
  Vanderburg A, Jontof-Hutter D, Rodriguez JE, Shporer A et~al. (2021) Tess
  discovery of a super-earth and three sub-neptunes hosted by the bright,
  sun-like star hd 108236.
\newblock \emph{The Astronomical Journal} 161(2): 85.

\bibitem[{Eisner et~al.(2020)Eisner, Barrag{\'a}n, Aigrain, Lintott, Miller,
  Zicher, Boyajian, Brice{\~n}o, Bryant, Christiansen
  et~al.}]{eisner2020planet}
Eisner N, Barrag{\'a}n O, Aigrain S, Lintott C, Miller G, Zicher N, Boyajian T,
  Brice{\~n}o C, Bryant E, Christiansen J et~al. (2020) Planet hunters tess i:
  Toi 813, a subgiant hosting a transiting saturn-sized planet on an 84-day
  orbit.
\newblock \emph{Monthly Notices of the Royal Astronomical Society} 494(1):
  750--763.

\bibitem[{Eisner et~al.(2022)Eisner, Johnston, Toonen, Frost, Janssens,
  Lintott, Aigrain, Sana, Abdul-Masih, Arellano-C{\'o}rdova
  et~al.}]{eisner2022planet}
Eisner NL, Johnston C, Toonen S, Frost AJ, Janssens S, Lintott CJ, Aigrain S,
  Sana H, Abdul-Masih M, Arellano-C{\'o}rdova KZ et~al. (2022) Planet hunters
  tess iv: a massive, compact hierarchical triple star system tic 470710327.
\newblock \emph{Monthly Notices of the Royal Astronomical Society} 511(4):
  4710--4723.

\bibitem[{Fausnaugh et~al.(2020)Fausnaugh, Burke, Ricker and
  Vanderspek}]{fausnaugh2020calibrated}
Fausnaugh MM, Burke CJ, Ricker GR and Vanderspek R (2020) Calibrated full-frame
  images for the tess quick look pipeline.
\newblock \emph{Research Notes of the AAS} 4(12): 251.

\bibitem[{Fischer et~al.(2012)Fischer, Schwamb, Schawinski, Lintott, Brewer,
  Giguere, Lynn, Parrish, Sartori, Simpson et~al.}]{fischer2012planet}
Fischer DA, Schwamb ME, Schawinski K, Lintott C, Brewer J, Giguere M, Lynn S,
  Parrish M, Sartori T, Simpson R et~al. (2012) Planet hunters: the first two
  planet candidates identified by the public using the kepler public archive
  data.
\newblock \emph{Monthly Notices of the Royal Astronomical Society} 419(4):
  2900--2911.

\bibitem[{Foreman-Mackey et~al.(2016)Foreman-Mackey, Morton, Hogg, Agol and
  Sch{\"o}lkopf}]{foreman2016population}
Foreman-Mackey D, Morton TD, Hogg DW, Agol E and Sch{\"o}lkopf B (2016) The
  population of long-period transiting exoplanets.
\newblock \emph{The Astronomical Journal} 152(6): 206.

\bibitem[{Gaidos et~al.(2019)Gaidos, Jacobs, LaCourse, Vanderburg, Rappaport,
  Berger, Pearce, Mann, Weiss, Fulton et~al.}]{gaidos2019planetesimals}
Gaidos E, Jacobs T, LaCourse D, Vanderburg A, Rappaport S, Berger T, Pearce L,
  Mann A, Weiss L, Fulton B et~al. (2019) Planetesimals around stars with tess
  (past)--i. transient dimming of a binary solar analogue at the end of the
  planet accretion era.
\newblock \emph{Monthly Notices of the Royal Astronomical Society} 488(4):
  4465--4476.

\bibitem[{Genova et~al.(2000)Genova, Egret, Bienaym{\'e}, Bonnarel, Dubois,
  Fernique, Jasniewicz, Lesteven, Monier, Ochsenbein et~al.}]{genova2000cds}
Genova F, Egret D, Bienaym{\'e} O, Bonnarel F, Dubois P, Fernique P, Jasniewicz
  G, Lesteven S, Monier R, Ochsenbein F et~al. (2000) The cds information
  hub-on--line services and links at the centre de donn{\'e}es astronomiques de
  strasbourg.
\newblock \emph{Astronomy and astrophysics supplement series} 143(1): 1--7.

\bibitem[{Gies et~al.(2013)Gies, Guo, Howell, Still, Boyajian, Hoekstra, Jek,
  LaCourse and Winarski}]{gies2013kic}
Gies DR, Guo Z, Howell SB, Still MD, Boyajian TS, Hoekstra AJ, Jek KJ, LaCourse
  D and Winarski T (2013) Kic 9406652: An unusual cataclysmic variable in the
  kepler field of view.
\newblock \emph{The Astrophysical Journal} 775(1): 64.

\bibitem[{Giles et~al.(2018)Giles, Osborn, Blanco-Cuaresma, Lovis, Bayliss,
  Eggenberger, Cameron, Kristiansen, Turner, Bouchy
  et~al.}]{giles2018transiting}
Giles H, Osborn H, Blanco-Cuaresma S, Lovis C, Bayliss D, Eggenberger P,
  Cameron AC, Kristiansen M, Turner O, Bouchy F et~al. (2018) Transiting planet
  candidate from k2 with the longest period.
\newblock \emph{Astronomy \& Astrophysics} 615: L13.

\bibitem[{Grindlay et~al.(2009)Grindlay, Tang, Simcoe, Laycock, Los, Mink,
  Doane and Champine}]{grindlay2009dasch}
Grindlay J, Tang S, Simcoe R, Laycock S, Los E, Mink D, Doane A and Champine G
  (2009) Dasch to measure (and preserve) the harvard plates: Opening the\~{}
  100-year time domain astronomy window.
\newblock In: \emph{Preserving Astronomy's Photographic Legacy: Current State
  and the Future of North American Astronomical Plates}, volume 410. p. 101.

\bibitem[{G{\"u}nther et~al.(2022)G{\"u}nther, Berardo, Ducrot, Murray,
  Stassun, Olah, Bouma, Rappaport, Winn, Feinstein et~al.}]{gunther2020complex}
G{\"u}nther MN, Berardo DA, Ducrot E, Murray CA, Stassun KG, Olah K, Bouma L,
  Rappaport S, Winn JN, Feinstein AD et~al. (2022) Complex modulation of
  rapidly rotating young m dwarfs: Adding pieces to the puzzle.
\newblock \emph{The astronomical journal} 163(4): 144.

\bibitem[{G{\"u}nther et~al.(2020)G{\"u}nther, Zhan, Seager, Rimmer, Ranjan,
  Stassun, Oelkers, Daylan, Newton, Kristiansen et~al.}]{gunther2020stellar}
G{\"u}nther MN, Zhan Z, Seager S, Rimmer PB, Ranjan S, Stassun KG, Oelkers RJ,
  Daylan T, Newton E, Kristiansen MH et~al. (2020) Stellar flares from the
  first tess data release: exploring a new sample of m dwarfs.
\newblock \emph{The Astronomical Journal} 159(2): 60.

\bibitem[{Han et~al.(2021)Han, Rappaport, Vanderburg, Tofflemire, Borkovits,
  Schwengeler, Zasche, Krolikowski, Muirhead, Kristiansen et~al.}]{han20212+}
Han E, Rappaport S, Vanderburg A, Tofflemire B, Borkovits T, Schwengeler H,
  Zasche P, Krolikowski D, Muirhead P, Kristiansen M et~al. (2021) A 2+ 1+ 1
  quadruple star system containing the most eccentric, low-mass, short-period,
  eclipsing binary known.
\newblock \emph{Monthly Notices of the Royal Astronomical Society} .

\bibitem[{Handler et~al.(2020)Handler, Kurtz, Rappaport, Saio, Fuller, Jones,
  Guo, Chowdhury, Sowicka, Ali{\c{c}}avu{\c{s}} et~al.}]{handler2020tidally}
Handler G, Kurtz DW, Rappaport S, Saio H, Fuller J, Jones D, Guo Z, Chowdhury
  S, Sowicka P, Ali{\c{c}}avu{\c{s}} FK et~al. (2020) Tidally trapped
  pulsations in a close binary star system discovered by tess.
\newblock \emph{Nature Astronomy} 4(7): 684--689.

\bibitem[{Hardegree-Ullman et~al.(2021)Hardegree-Ullman, Christiansen, Ciardi,
  Crossfield, Dressing, Livingston, Volk, Agol, Barclay, Barentsen
  et~al.}]{hardegree2021k2}
Hardegree-Ullman KK, Christiansen JL, Ciardi DR, Crossfield IJ, Dressing CD,
  Livingston JH, Volk K, Agol E, Barclay T, Barentsen G et~al. (2021) K2-138 g:
  Spitzer spots a sixth planet for the citizen science system.
\newblock \emph{The Astronomical Journal} 161(5): 219.

\bibitem[{Hartman(2012)}]{hartman2012vartools}
Hartman J (2012) Vartools: Light curve analysis program.
\newblock \emph{Astrophysics Source Code Library} : ascl--1208.

\bibitem[{Hitchcock et~al.(2019)Hitchcock, Fossey and
  Savini}]{hitchcock2019non}
Hitchcock J, Fossey SJ and Savini G (2019) Non-gray, month-long brightening of
  kic 8462852 in the immediate aftermath of a deep dip.
\newblock \emph{Publications of the Astronomical Society of the Pacific}
  131(1002): 084204.

\bibitem[{Howell et~al.(2014)Howell, Sobeck, Haas, Still, Barclay, Mullally,
  Troeltzsch, Aigrain, Bryson, Caldwell et~al.}]{howell2014k2}
Howell SB, Sobeck C, Haas M, Still M, Barclay T, Mullally F, Troeltzsch J,
  Aigrain S, Bryson ST, Caldwell D et~al. (2014) The k2 mission:
  characterization and early results.
\newblock \emph{Publications of the Astronomical Society of the Pacific}
  126(938): 398.

\bibitem[{Huang et~al.(2020{\natexlab{a}})Huang, Vanderburg, P{\'a}l, Sha, Yu,
  Fong, Fausnaugh, Shporer, Guerrero, Vanderspek et~al.}]{huang2020photometry1}
Huang CX, Vanderburg A, P{\'a}l A, Sha L, Yu L, Fong W, Fausnaugh M, Shporer A,
  Guerrero N, Vanderspek R et~al. (2020{\natexlab{a}}) Photometry of 10 million
  stars from the first two years of tess full frame images: Part i.
\newblock \emph{Research Notes of the AAS} 4(11): 204.

\bibitem[{Huang et~al.(2020{\natexlab{b}})Huang, Vanderburg, P{\'a}l, Sha, Yu,
  Fong, Fausnaugh, Shporer, Guerrero, Vanderspek et~al.}]{huang2020photometry2}
Huang CX, Vanderburg A, P{\'a}l A, Sha L, Yu L, Fong W, Fausnaugh M, Shporer A,
  Guerrero N, Vanderspek R et~al. (2020{\natexlab{b}}) Photometry of 10 million
  stars from the first two years of tess full frame images: Part ii.
\newblock \emph{Research Notes of the AAS} 4(11): 206.

\bibitem[{Ilin et~al.(2021)Ilin, Schmidt, Poppenh{\"a}ger, Davenport,
  Kristiansen and Omohundro}]{ilin2021flares}
Ilin E, Schmidt SJ, Poppenh{\"a}ger K, Davenport JR, Kristiansen MH and
  Omohundro M (2021) Flares in open clusters with k2-ii. pleiades, hyades,
  praesepe, ruprecht 147, and m 67.
\newblock \emph{Astronomy \& Astrophysics} 645: A42.

\bibitem[{Jayaraman et~al.(2022)Jayaraman, Hubrig, Holdsworth, Sch{\"o}ller,
  J{\"a}rvinen, Kurtz, Gagliano and Ricker}]{jayaraman2022could}
Jayaraman R, Hubrig S, Holdsworth DL, Sch{\"o}ller M, J{\"a}rvinen S, Kurtz DW,
  Gagliano R and Ricker GR (2022) Could the magnetic star hd 135348 possess a
  rigidly rotating magnetosphere?
\newblock \emph{The Astrophysical Journal Letters} 924(1): L10.

\bibitem[{Jenkins et~al.(2002)Jenkins, Caldwell and Borucki}]{jenkins2002some}
Jenkins JM, Caldwell DA and Borucki WJ (2002) Some tests to establish
  confidence in planets discovered by transit photometry.
\newblock \emph{The Astrophysical Journal} 564(1): 495.

\bibitem[{Jenkins et~al.(2010{\natexlab{a}})Jenkins, Caldwell, Chandrasekaran,
  Twicken, Bryson, Quintana, Clarke, Li, Allen, Tenenbaum
  et~al.}]{jenkins2010overview}
Jenkins JM, Caldwell DA, Chandrasekaran H, Twicken JD, Bryson ST, Quintana EV,
  Clarke BD, Li J, Allen C, Tenenbaum P et~al. (2010{\natexlab{a}}) Overview of
  the kepler science processing pipeline.
\newblock \emph{The Astrophysical Journal Letters} 713(2): L87.

\bibitem[{Jenkins et~al.(2010{\natexlab{b}})Jenkins, Chandrasekaran, McCauliff,
  Caldwell, Tenenbaum, Li, Klaus, Cote and Middour}]{jenkins2010transiting}
Jenkins JM, Chandrasekaran H, McCauliff SD, Caldwell DA, Tenenbaum P, Li J,
  Klaus TC, Cote MT and Middour C (2010{\natexlab{b}}) Transiting planet search
  in the kepler pipeline.
\newblock In: \emph{Software and Cyberinfrastructure for Astronomy}, volume
  7740. International Society for Optics and Photonics, p. 77400D.

\bibitem[{Kane et~al.(2020)Kane, Yal{\c{c}}{\i}nkaya, Osborn, Dalba, Nielsen,
  Vanderburg, Mo{\v{c}}nik, Hinkel, Ostberg, Esmer et~al.}]{kane2020transits}
Kane SR, Yal{\c{c}}{\i}nkaya S, Osborn HP, Dalba PA, Nielsen LD, Vanderburg A,
  Mo{\v{c}}nik T, Hinkel NR, Ostberg C, Esmer EM et~al. (2020) Transits of
  known planets orbiting a naked-eye star.
\newblock \emph{The Astronomical Journal} 160(3): 129.

\bibitem[{Kennedy et~al.(2019)Kennedy, Hope, Hodgkin and
  Wyatt}]{kennedy2019automated}
Kennedy GM, Hope G, Hodgkin ST and Wyatt MC (2019) An automated search for
  transiting exocomets.
\newblock \emph{Monthly Notices of the Royal Astronomical Society} 482(4):
  5587--5596.

\bibitem[{Kipping et~al.(2015)Kipping, Schmitt, Huang, Torres, Nesvorn{\`y},
  Buchhave, Hartman and Bakos}]{kipping2015hunt}
Kipping DM, Schmitt AR, Huang X, Torres G, Nesvorn{\`y} D, Buchhave LA, Hartman
  J and Bakos G{\'A} (2015) The hunt for exomoons with kepler (hek). v. a
  survey of 41 planetary candidates for exomoons.
\newblock \emph{The Astrophysical Journal} 813(1): 14.

\bibitem[{Kirk et~al.(2016)Kirk, Conroy, Pr{\v{s}}a, Abdul-Masih, Kochoska,
  Matijevi{\v{C}}, Hambleton, Barclay, Bloemen, Boyajian
  et~al.}]{kirk2016kepler}
Kirk B, Conroy K, Pr{\v{s}}a A, Abdul-Masih M, Kochoska A, Matijevi{\v{C}} G,
  Hambleton K, Barclay T, Bloemen S, Boyajian T et~al. (2016) Kepler eclipsing
  binary stars. vii. the catalog of eclipsing binaries found in the entire
  kepler data set.
\newblock \emph{The Astronomical Journal} 151(3): 68.

\bibitem[{Koch et~al.(2010)Koch, Borucki, Basri, Batalha, Brown, Caldwell,
  Christensen-Dalsgaard, Cochran, DeVore, Dunham et~al.}]{koch2010kepler}
Koch DG, Borucki WJ, Basri G, Batalha NM, Brown TM, Caldwell D,
  Christensen-Dalsgaard J, Cochran WD, DeVore E, Dunham EW et~al. (2010) Kepler
  mission design, realized photometric performance, and early science.
\newblock \emph{The Astrophysical Journal Letters} 713(2): L79.

\bibitem[{Kostov et~al.(2022)Kostov, Powell, Rappaport, Borkovits, Gagliano,
  Jacobs, Kristiansen, LaCourse, Omohundro, Orosz et~al.}]{kostov202297}
Kostov VB, Powell BP, Rappaport SA, Borkovits T, Gagliano R, Jacobs TL,
  Kristiansen MH, LaCourse DM, Omohundro M, Orosz J et~al. (2022) 97 eclipsing
  quadruple star candidates discovered in tess full frame images.
\newblock \emph{arXiv preprint arXiv:2202.05790} .

\bibitem[{Kurtz et~al.(2020)Kurtz, Handler, Rappaport, Saio, Fuller, Jacobs,
  Schmitt, Jones, Vanderburg, LaCourse et~al.}]{kurtz2020single}
Kurtz DW, Handler G, Rappaport S, Saio H, Fuller J, Jacobs T, Schmitt A, Jones
  D, Vanderburg A, LaCourse D et~al. (2020) The single-sided pulsator co
  camelopardalis.
\newblock \emph{Monthly Notices of the Royal Astronomical Society} 494(4):
  5118--5133.

\bibitem[{LaCourse and Jacobs(2018)}]{lacourse2018single2}
LaCourse DM and Jacobs TL (2018) Single transits and eclipses observed by k2.
\newblock \emph{Research Notes of the AAS} 2(1): 28.

\bibitem[{LaCourse et~al.(2015)LaCourse, Jek, Jacobs, Winarski, Boyajian,
  Rappaport, Sanchis-Ojeda, Conroy, Nelson, Barclay
  et~al.}]{lacourse2015kepler}
LaCourse DM, Jek KJ, Jacobs TL, Winarski T, Boyajian TS, Rappaport SA,
  Sanchis-Ojeda R, Conroy KE, Nelson L, Barclay T et~al. (2015) Kepler
  eclipsing binary stars--vi. identification of eclipsing binaries in the k2
  campaign 0 data set.
\newblock \emph{Monthly Notices of the Royal Astronomical Society} 452(4):
  3561--3592.

\bibitem[{Lee et~al.(2018)Lee, Hong and Kristiansen}]{lee2018eclipsing}
Lee JW, Hong K and Kristiansen MH (2018) The eclipsing $\delta$ scuti star epic
  245932119.
\newblock \emph{The Astronomical Journal} 157(1): 17.

\bibitem[{Lee et~al.(2020)Lee, Hong and Kristiansen}]{lee2020tess}
Lee JW, Hong K and Kristiansen MH (2020) Tess photometry of the eclipsing
  $\delta$ scuti star ai hydrae.
\newblock \emph{Publications of the Astronomical Society of Japan} 72(3): 37.

\bibitem[{Lee et~al.(2019)Lee, Kristiansen and Hong}]{lee2019pulsating}
Lee JW, Kristiansen MH and Hong K (2019) The pulsating eclipsing binary tic
  309658221 in a 7.59-day orbit.
\newblock \emph{The Astronomical Journal} 157(6): 223.

\bibitem[{Leleu et~al.(2021)Leleu, Alibert, Hara, Hooton, Wilson, Robutel,
  Delisle, Laskar, Hoyer, Lovis et~al.}]{leleu2021six}
Leleu A, Alibert Y, Hara N, Hooton M, Wilson T, Robutel P, Delisle JB, Laskar
  J, Hoyer S, Lovis C et~al. (2021) Six transiting planets and a chain of
  laplace resonances in toi-178.
\newblock \emph{Astronomy \& Astrophysics} 649: A26.

\bibitem[{{Lightkurve Collaboration}:~Cardoso et~al.(2018){Lightkurve
  Collaboration}:~Cardoso, Hedges, Gully-Santiago, Saunders, Cody, Barclay,
  Hall, Sagear, Turtelboom, Zhang et~al.}]{cardoso2018lightkurve}
{Lightkurve Collaboration}:~Cardoso JVdM, Hedges C, Gully-Santiago M, Saunders
  N, Cody AM, Barclay T, Hall O, Sagear S, Turtelboom E, Zhang J et~al. (2018)
  Lightkurve: Kepler and tess time series analysis in python.
\newblock \emph{Astrophysics Source Code Library} : ascl--1812.

\bibitem[{Lintott(2020)}]{lintott2020citizen}
Lintott C (2020) Citizen science: The past 200 years.
\newblock \emph{Astronomy \& Geophysics} 61(2): 2--20.

\bibitem[{Lintott et~al.(2008)Lintott, Schawinski, Slosar, Land, Bamford,
  Thomas, Raddick, Nichol, Szalay, Andreescu et~al.}]{lintott2008galaxy}
Lintott CJ, Schawinski K, Slosar A, Land K, Bamford S, Thomas D, Raddick MJ,
  Nichol RC, Szalay A, Andreescu D et~al. (2008) Galaxy zoo: morphologies
  derived from visual inspection of galaxies from the sloan digital sky survey.
\newblock \emph{Monthly Notices of the Royal Astronomical Society} 389(3):
  1179--1189.

\bibitem[{Lintott et~al.(2013)Lintott, Schwamb, Barclay, Sharzer, Fischer,
  Brewer, Giguere, Lynn, Parrish, Batalha et~al.}]{lintott2013planet}
Lintott CJ, Schwamb ME, Barclay T, Sharzer C, Fischer DA, Brewer J, Giguere M,
  Lynn S, Parrish M, Batalha N et~al. (2013) Planet hunters: New kepler planet
  candidates from analysis of quarter 2.
\newblock \emph{The Astronomical Journal} 145(6): 151.

\bibitem[{Malavolta et~al.(2018)Malavolta, Mayo, Louden, Rajpaul, Bonomo,
  Buchhave, Kreidberg, Kristiansen, Lopez-Morales, Mortier
  et~al.}]{malavolta2018ultra}
Malavolta L, Mayo AW, Louden T, Rajpaul VM, Bonomo AS, Buchhave LA, Kreidberg
  L, Kristiansen MH, Lopez-Morales M, Mortier A et~al. (2018) An ultra-short
  period rocky super-earth with a secondary eclipse and a neptune-like
  companion around k2-141.
\newblock \emph{The Astronomical Journal} 155(3): 107.

\bibitem[{Mann et~al.(2016)Mann, Gaidos, Mace, Johnson, Bowler, LaCourse,
  Jacobs, Vanderburg, Kraus, Kaplan et~al.}]{mann2016zodiacal}
Mann AW, Gaidos E, Mace GN, Johnson MC, Bowler BP, LaCourse D, Jacobs TL,
  Vanderburg A, Kraus AL, Kaplan KF et~al. (2016) Zodiacal exoplanets in time
  (zeit). i. a neptune-sized planet orbiting an m4. 5 dwarf in the hyades star
  cluster.
\newblock \emph{The Astrophysical Journal} 818(1): 46.

\bibitem[{Mann et~al.(2020)Mann, Johnson, Vanderburg, Kraus, Rizzuto, Wood,
  Bush, Rockcliffe, Newton, Latham et~al.}]{mann2020tess}
Mann AW, Johnson MC, Vanderburg A, Kraus AL, Rizzuto AC, Wood ML, Bush JL,
  Rockcliffe K, Newton ER, Latham DW et~al. (2020) Tess hunt for young and
  maturing exoplanets (thyme). iii. a two-planet system in the 400 myr ursa
  major group.
\newblock \emph{The Astronomical Journal} 160(4): 179.

\bibitem[{Morton(2015)}]{morton2015vespa}
Morton TD (2015) Vespa: False positive probabilities calculator.
\newblock \emph{Astrophysics Source Code Library} : ascl--1503.

\bibitem[{Nardiello et~al.(2019)Nardiello, Borsato, Piotto, Colombo,
  Manthopoulou, Bedin, Granata, Lacedelli, Libralato, Malavolta
  et~al.}]{nardiello2019psf}
Nardiello D, Borsato L, Piotto G, Colombo L, Manthopoulou E, Bedin L, Granata
  V, Lacedelli G, Libralato M, Malavolta L et~al. (2019) A psf-based approach
  to tess high quality data of stellar clusters (pathos)--i. search for
  exoplanets and variable stars in the field of 47 tuc.
\newblock \emph{Monthly Notices of the Royal Astronomical Society} 490(3):
  3806--3823.

\bibitem[{Newberg et~al.(2013)Newberg, Newby, Desell, Magdon-Ismail, Szymanski
  and Varela}]{newberg2013milkyway}
Newberg HJ, Newby M, Desell T, Magdon-Ismail M, Szymanski B and Varela C (2013)
  Milkyway@ home: Harnessing volunteer computers to constrain dark matter in
  the milky way.
\newblock \emph{Proceedings of the International Astronomical Union} 9(S298):
  98--104.

\bibitem[{Oelkers and Stassun(2018)}]{oelkers2018precision}
Oelkers RJ and Stassun KG (2018) Precision light curves from tess full-frame
  images: A different imaging approach.
\newblock \emph{The Astronomical Journal} 156(3): 132.

\bibitem[{Ol{\'a}h et~al.(2018)Ol{\'a}h, Rappaport, Borkovits, Jacobs, Latham,
  Bieryla, B{\'\i}r{\'o}, Bartus, K{\H{o}}v{\'a}ri, Vida
  et~al.}]{olah2018eclipsing}
Ol{\'a}h K, Rappaport S, Borkovits T, Jacobs T, Latham D, Bieryla A,
  B{\'\i}r{\'o} I, Bartus J, K{\H{o}}v{\'a}ri Z, Vida K et~al. (2018) Eclipsing
  spotted giant star with k2 and historical photometry.
\newblock \emph{Astronomy \& Astrophysics} 620: A189.

\bibitem[{Ol{\'a}h et~al.(2020)Ol{\'a}h, Rappaport, Derekas and
  Vanderburg}]{olah2020importance}
Ol{\'a}h K, Rappaport S, Derekas A and Vanderburg A (2020) The importance of
  studying active giant stars in eclipsing binaries--and the role of citizen
  scientists in finding them.
\newblock \emph{Contrib. Astron. Obs. Skalnat{\'e} Pleso} 50: 390--394.

\bibitem[{Olmschenk et~al.(2021)Olmschenk, Silva, Rau, Barry, Kruse,
  Cacciapuoti, Kostov, Powell, Wyrwas, Schnittman
  et~al.}]{olmschenk2021identifying}
Olmschenk G, Silva SI, Rau G, Barry RK, Kruse E, Cacciapuoti L, Kostov V,
  Powell BP, Wyrwas E, Schnittman JD et~al. (2021) Identifying planetary
  transit candidates in tess full-frame image light curves via convolutional
  neural networks.
\newblock \emph{The Astronomical Journal} 161(6): 273.

\bibitem[{Osborn(2017)}]{osborn2017long}
Osborn H (2017) Long-period exoplanets from photometric transit surveys.
\newblock \emph{Ph. D. Thesis} .

\bibitem[{Osborn(2021)}]{osbornmonotools}
Osborn H (2021) Monotools – a python package for planets of uncertain period.
\newblock \emph{The Journal of Open Source Software} : x.

\bibitem[{Osborn et~al.(2016)Osborn, Armstrong, Brown, McCormac, Doyle, Louden,
  Kirk, Spake, Lam, Walker et~al.}]{osborn2016single}
Osborn H, Armstrong D, Brown D, McCormac J, Doyle A, Louden T, Kirk J, Spake J,
  Lam K, Walker S et~al. (2016) Single transit candidates from k2: detection
  and period estimation.
\newblock \emph{Monthly Notices of the Royal Astronomical Society} 457(3):
  2273--2286.

\bibitem[{Pence and Chai(2012)}]{pence2012fv}
Pence W and Chai P (2012) Fv: Interactive fits file editor.
\newblock \emph{Astrophysics Source Code Library} : ascl--1205.

\bibitem[{Pojmanski(1997)}]{pojmanski1997all2}
Pojmanski G (1997) The all sky automated survey.
\newblock \emph{Acta Astronomica} 47: 467--481.

\bibitem[{Powell et~al.(2021{\natexlab{a}})Powell, Kostov, Rappaport,
  Borkovits, Zasche, Tokovinin, Kruse, Latham, Montet, Jensen
  et~al.}]{powell2021tic}
Powell BP, Kostov VB, Rappaport SA, Borkovits T, Zasche P, Tokovinin A, Kruse
  E, Latham DW, Montet BT, Jensen EL et~al. (2021{\natexlab{a}}) Tic 168789840:
  A sextuply eclipsing sextuple star system.
\newblock \emph{The Astronomical Journal} 161(4): 162.

\bibitem[{Powell et~al.(2021{\natexlab{b}})Powell, Kostov, Rappaport,
  Tokovinin, Shporer, Collins, Corbett, Borkovits, Gary, Chiang
  et~al.}]{powell2021mysterious}
Powell BP, Kostov VB, Rappaport SA, Tokovinin A, Shporer A, Collins KA, Corbett
  H, Borkovits T, Gary BL, Chiang E et~al. (2021{\natexlab{b}}) Mysterious
  dust-emitting object orbiting tic 400799224.
\newblock \emph{The Astronomical Journal} 162(6): 299.

\bibitem[{Pr{\v{s}}a et~al.(2022)Pr{\v{s}}a, Kochoska, Conroy, Eisner, Hey,
  IJspeert, Kruse, Fleming, Johnston, Kristiansen et~al.}]{prvsa2022tess}
Pr{\v{s}}a A, Kochoska A, Conroy KE, Eisner N, Hey DR, IJspeert L, Kruse E,
  Fleming SW, Johnston C, Kristiansen MH et~al. (2022) Tess eclipsing binary
  stars. i. short-cadence observations of 4584 eclipsing binaries in sectors
  1--26.
\newblock \emph{The Astrophysical Journal Supplement Series} 258(1): 16.

\bibitem[{Quinn et~al.(2019)Quinn, Becker, Rodriguez, Hadden, Huang, Morton,
  Adams, Armstrong, Eastman, Horner et~al.}]{quinn2019near}
Quinn SN, Becker JC, Rodriguez JE, Hadden S, Huang CX, Morton TD, Adams FC,
  Armstrong D, Eastman JD, Horner J et~al. (2019) Near-resonance in a system of
  sub-neptunes from tess.
\newblock \emph{The Astronomical Journal} 158(5): 177.

\bibitem[{Quinn et~al.(2021)Quinn, Rappaport, Vanderburg, Eastman, Nelson,
  Jacobs, LaCourse, Schmitt, Berlind, Calkins et~al.}]{quinn2021long}
Quinn SN, Rappaport S, Vanderburg A, Eastman JD, Nelson LA, Jacobs TL, LaCourse
  DM, Schmitt AR, Berlind P, Calkins ML et~al. (2021) A long-period substellar
  object exhibiting a single transit in kepler.
\newblock \emph{arXiv preprint arXiv:2107.00027} .

\bibitem[{Rappaport et~al.(2022)Rappaport, Borkovits, Gagliano, Jacobs, Kostov,
  Powell, Terentev, Omohundro, Torres, Vanderburg et~al.}]{rappaport2022six}
Rappaport S, Borkovits T, Gagliano R, Jacobs T, Kostov V, Powell B, Terentev I,
  Omohundro M, Torres G, Vanderburg A et~al. (2022) Six new compact triply
  eclipsing triples found with tess.
\newblock \emph{Monthly Notices of the Royal Astronomical Society} .

\bibitem[{Rappaport et~al.(2016)Rappaport, Lehmann, Kalomeni, Borkovits,
  Latham, Bieryla, Ngo, Mawet, Howell, Horch et~al.}]{rappaport2016quintuple}
Rappaport S, Lehmann H, Kalomeni B, Borkovits T, Latham D, Bieryla A, Ngo H,
  Mawet D, Howell S, Horch E et~al. (2016) A quintuple star system containing
  two eclipsing binaries.
\newblock \emph{Monthly Notices of the Royal Astronomical Society} 462(2):
  1812--1825.

\bibitem[{Rappaport et~al.(2017)Rappaport, Vanderburg, Borkovits, Kalomeni,
  Halpern, Ngo, Mace, Fulton, Howard, Isaacson et~al.}]{rappaport2017epic}
Rappaport S, Vanderburg A, Borkovits T, Kalomeni B, Halpern J, Ngo H, Mace G,
  Fulton B, Howard A, Isaacson H et~al. (2017) Epic 220204960: a quadruple star
  system containing two strongly interacting eclipsing binaries.
\newblock \emph{Monthly Notices of the Royal Astronomical Society} 467(2):
  2160--2179.

\bibitem[{Rappaport et~al.(2018)Rappaport, Vanderburg, Jacobs, LaCourse,
  Jenkins, Kraus, Rizzuto, Latham, Bieryla, Lazarevic
  et~al.}]{rappaport2018likely}
Rappaport S, Vanderburg A, Jacobs T, LaCourse D, Jenkins J, Kraus A, Rizzuto A,
  Latham D, Bieryla A, Lazarevic M et~al. (2018) Likely transiting exocomets
  detected by kepler.
\newblock \emph{Monthly Notices of the Royal Astronomical Society} 474(2):
  1453--1468.

\bibitem[{Rappaport et~al.(2019{\natexlab{a}})Rappaport, Vanderburg,
  Kristiansen, Omohundro, Schwengeler, Terentev, Dai, Masuda, Jacobs, LaCourse
  et~al.}]{rappaport2019random}
Rappaport S, Vanderburg A, Kristiansen M, Omohundro M, Schwengeler H, Terentev
  I, Dai F, Masuda K, Jacobs T, LaCourse D et~al. (2019{\natexlab{a}}) The
  random transiter--epic 249706694/hd 139139.
\newblock \emph{Monthly Notices of the Royal Astronomical Society} 488(2):
  2455--2465.

\bibitem[{Rappaport et~al.(2019{\natexlab{b}})Rappaport, Zhou, Vanderburg,
  Mann, Kristiansen, Ol{\'a}h, Jacobs, Newton, Omohundro, LaCourse
  et~al.}]{rappaport2019deep}
Rappaport S, Zhou G, Vanderburg A, Mann A, Kristiansen M, Ol{\'a}h K, Jacobs T,
  Newton E, Omohundro M, LaCourse D et~al. (2019{\natexlab{b}}) Deep long
  asymmetric occultation in epic 204376071.
\newblock \emph{Monthly Notices of the Royal Astronomical Society} 485(2):
  2681--2693.

\bibitem[{Ricker et~al.(2014)Ricker, Winn, Vanderspek, Latham, Bakos, Bean,
  Berta-Thompson, Brown, Buchhave, Butler et~al.}]{ricker2014transiting}
Ricker GR, Winn JN, Vanderspek R, Latham DW, Bakos G{\'A}, Bean JL,
  Berta-Thompson ZK, Brown TM, Buchhave L, Butler NR et~al. (2014) Transiting
  exoplanet survey satellite.
\newblock \emph{Journal of Astronomical Telescopes, Instruments, and Systems}
  1(1): 014003.

\bibitem[{Rodriguez et~al.(2018)Rodriguez, Becker, Eastman, Hadden, Vanderburg,
  Khain, Quinn, Mayo, Dressing, Schlieder et~al.}]{rodriguez2018compact}
Rodriguez JE, Becker JC, Eastman JD, Hadden S, Vanderburg A, Khain T, Quinn SN,
  Mayo A, Dressing CD, Schlieder JE et~al. (2018) A compact multi-planet system
  with a significantly misaligned ultra short period planet.
\newblock \emph{The Astronomical Journal} 156(5): 245.

\bibitem[{Schmitt and Vanderburg(2021)}]{schmitt2021lctools}
Schmitt A and Vanderburg A (2021) Lctools ii: The quickfind method for finding
  signals and associated ttvs in light curves from nasa space missions.
\newblock \emph{arXiv preprint arXiv:2103.10285} .

\bibitem[{Schmitt et~al.(2019)Schmitt, Hartman and
  Kipping}]{schmitt2019lctools}
Schmitt AR, Hartman JD and Kipping DM (2019) Lctools: a windows-based software
  system for finding and recording signals in lightcurves from nasa space
  missions.
\newblock \emph{arXiv preprint arXiv:1910.08034} .

\bibitem[{Schmitt et~al.(2014{\natexlab{a}})Schmitt, Agol, Deck, Rogers, Gazak,
  Fischer, Wang, Holman, Jek, Margossian et~al.}]{schmitt2014planet2}
Schmitt JR, Agol E, Deck KM, Rogers LA, Gazak JZ, Fischer DA, Wang J, Holman
  MJ, Jek KJ, Margossian C et~al. (2014{\natexlab{a}}) Planet hunters. vii.
  discovery of a new low-mass, low-density planet (ph3 c) orbiting kepler-289
  with mass measurements of two additional planets (ph3 b and d).
\newblock \emph{The Astrophysical Journal} 795(2): 167.

\bibitem[{Schmitt et~al.(2017)Schmitt, Jenkins and Fischer}]{schmitt2017search}
Schmitt JR, Jenkins JM and Fischer DA (2017) A search for lost planets in the
  kepler multi-planet systems and the discovery of the long-period,
  neptune-sized exoplanet kepler-150 f.
\newblock \emph{The Astronomical Journal} 153(4): 180.

\bibitem[{Schmitt et~al.(2016)Schmitt, Tokovinin, Wang, Fischer, Kristiansen,
  LaCourse, Gagliano, Tan, Schwengeler, Omohundro et~al.}]{schmitt2016planet}
Schmitt JR, Tokovinin A, Wang J, Fischer DA, Kristiansen MH, LaCourse DM,
  Gagliano R, Tan AJV, Schwengeler HM, Omohundro MR et~al. (2016) Planet
  hunters. x. searching for nearby neighbors of 75 planet and eclipsing binary
  candidates from the k2 kepler extended mission.
\newblock \emph{The Astronomical Journal} 151(6): 159.

\bibitem[{Schmitt et~al.(2014{\natexlab{b}})Schmitt, Wang, Fischer, Jek,
  Moriarty, Boyajian, Schwamb, Lintott, Lynn, Smith et~al.}]{schmitt2014planet}
Schmitt JR, Wang J, Fischer DA, Jek KJ, Moriarty JC, Boyajian TS, Schwamb ME,
  Lintott C, Lynn S, Smith AM et~al. (2014{\natexlab{b}}) Planet hunters. vi.
  an independent characterization of koi-351 and several long period planet
  candidates from the kepler archival data.
\newblock \emph{The Astronomical Journal} 148(2): 28.

\bibitem[{Schmitt et~al.(2015)Schmitt, Wang, Fischer, Jek, Moriarty, Boyajian,
  Schwamb, Lintott, Lynn, Smith et~al.}]{schmitt2015erratum}
Schmitt JR, Wang J, Fischer DA, Jek KJ, Moriarty JC, Boyajian TS, Schwamb ME,
  Lintott C, Lynn S, Smith AM et~al. (2015) Erratum:“planet hunters. vi. an
  independent characterization of koi-351 and several long period planet
  candidates from the kepler archival data”(2014, aj, 148, 28).
\newblock \emph{The Astronomical Journal} 150(1): 38.

\bibitem[{Schwamb et~al.(2013)Schwamb, Orosz, Carter, Welsh, Fischer, Torres,
  Howard, Crepp, Keel, Lintott et~al.}]{schwamb2013planet}
Schwamb ME, Orosz JA, Carter JA, Welsh WF, Fischer DA, Torres G, Howard AW,
  Crepp JR, Keel WC, Lintott CJ et~al. (2013) Planet hunters: A transiting
  circumbinary planet in a quadruple star system.
\newblock \emph{The Astrophysical Journal} 768(2): 127.

\bibitem[{Shappee et~al.(2014)Shappee, Prieto, Stanek, Kochanek, Holoien,
  Jencson, Basu, Beacom, Szczygiel, Pojmanski et~al.}]{shappee2014all}
Shappee B, Prieto J, Stanek K, Kochanek C, Holoien T, Jencson J, Basu U, Beacom
  J, Szczygiel D, Pojmanski G et~al. (2014) All sky automated survey for
  supernovae (asas-sn or" assassin").
\newblock In: \emph{American Astronomical Society Meeting Abstracts\# 223},
  volume 223. pp. 236--03.

\bibitem[{Stassun et~al.(2018)Stassun, Oelkers, Pepper, Paegert, De~Lee,
  Torres, Latham, Charpinet, Dressing, Huber et~al.}]{stassun2018tess}
Stassun KG, Oelkers RJ, Pepper J, Paegert M, De~Lee N, Torres G, Latham DW,
  Charpinet S, Dressing CD, Huber D et~al. (2018) The tess input catalog and
  candidate target list.
\newblock \emph{The Astronomical Journal} 156(3): 102.

\bibitem[{Still and Barclay(2012)}]{still2012pyke}
Still M and Barclay T (2012) Pyke: Reduction and analysis of kepler simple
  aperture photometry data.
\newblock \emph{Astrophysics Source Code Library} : ascl--1208.

\bibitem[{Taylor(2011)}]{taylor2011topcat}
Taylor M (2011) Topcat: tool for operations on catalogues and tables.
\newblock \emph{Astrophysics Source Code Library} 1(S 01010).

\bibitem[{Tonry et~al.(2018)Tonry, Denneau, Heinze, Stalder, Smith, Smartt,
  Stubbs, Weiland and Rest}]{tonry2018atlas}
Tonry J, Denneau L, Heinze A, Stalder B, Smith K, Smartt S, Stubbs C, Weiland H
  and Rest A (2018) Atlas: a high-cadence all-sky survey system.
\newblock \emph{Publications of the Astronomical Society of the Pacific}
  130(988): 064505.

\bibitem[{Uehara et~al.(2016)Uehara, Kawahara, Masuda, Yamada and
  Aizawa}]{uehara2016transiting}
Uehara S, Kawahara H, Masuda K, Yamada S and Aizawa M (2016) Transiting planet
  candidates beyond the snow line detected by visual inspection of 7557 kepler
  objects of interest.
\newblock \emph{The Astrophysical Journal} 822(1): 2.

\bibitem[{Vanderburg et~al.(2016)Vanderburg, Becker, Kristiansen, Bieryla,
  Duev, Jensen-Clem, Morton, Latham, Adams, Baranec
  et~al.}]{vanderburg2016five}
Vanderburg A, Becker JC, Kristiansen MH, Bieryla A, Duev DA, Jensen-Clem R,
  Morton TD, Latham DW, Adams FC, Baranec C et~al. (2016) Five planets
  transiting a ninth magnitude star.
\newblock \emph{The Astrophysical Journal Letters} 827(1): L10.

\bibitem[{Vanderburg and Johnson(2014)}]{vanderburg2014technique}
Vanderburg A and Johnson JA (2014) A technique for extracting highly precise
  photometry for the two-wheeled kepler mission.
\newblock \emph{Publications of the Astronomical Society of the Pacific}
  126(944): 948.

\bibitem[{Venner et~al.(2021)Venner, Vanderburg and Pearce}]{venner2021true}
Venner A, Vanderburg A and Pearce LA (2021) True masses of the long-period
  companions to hd 92987 and hd 221420 from hipparcos--gaia astrometry.
\newblock \emph{The Astronomical Journal} 162(1): 12.

\bibitem[{Villanueva~Jr et~al.(2019)Villanueva~Jr, Dragomir and
  Gaudi}]{villanueva2019estimate}
Villanueva~Jr S, Dragomir D and Gaudi BS (2019) An estimate of the yield of
  single-transit planetary events from the transiting exoplanet survey
  satellite.
\newblock \emph{The Astronomical Journal} 157(2): 84.

\bibitem[{Wang et~al.(2013{\natexlab{a}})Wang, Fischer, Barclay, Boyajian,
  Crepp, Schwamb, Lintott, Jek, Smith, Parrish et~al.}]{wang2013planet2}
Wang J, Fischer DA, Barclay T, Boyajian TS, Crepp JR, Schwamb ME, Lintott C,
  Jek KJ, Smith AM, Parrish M et~al. (2013{\natexlab{a}}) Erratum: Planet
  hunters. v. a confirmed jupiter-size planet in the habitable zone and 42
  planet candidates from the kepler archive data (2013, {ApJ}, 776, 10).
\newblock \emph{The Astrophysical Journal} 778(1): 84.

\bibitem[{Wang et~al.(2013{\natexlab{b}})Wang, Fischer, Barclay, Boyajian,
  Crepp, Schwamb, Lintott, Jek, Smith, Parrish et~al.}]{wang2013planet}
Wang J, Fischer DA, Barclay T, Boyajian TS, Crepp JR, Schwamb ME, Lintott C,
  Jek KJ, Smith AM, Parrish M et~al. (2013{\natexlab{b}}) Planet hunters. v. a
  confirmed jupiter-size planet in the habitable zone and 42 planet candidates
  from the kepler archive data.
\newblock \emph{The Astrophysical Journal} 776(1): 10.

\bibitem[{Wang et~al.(2015)Wang, Fischer, Barclay, Picard, Ma, Bowler, Schmitt,
  Boyajian, Jek, LaCourse et~al.}]{wang2015planet}
Wang J, Fischer DA, Barclay T, Picard A, Ma B, Bowler BP, Schmitt JR, Boyajian
  TS, Jek KJ, LaCourse D et~al. (2015) Planet hunters. viii. characterization
  of 41 long-period exoplanet candidates from kepler archival data.
\newblock \emph{The Astrophysical Journal} 815(2): 127.

\bibitem[{Wenger et~al.(2000)Wenger, Ochsenbein, Egret, Dubois, Bonnarel,
  Borde, Genova, Jasniewicz, Lalo{\"e}, Lesteven et~al.}]{wenger2000simbad}
Wenger M, Ochsenbein F, Egret D, Dubois P, Bonnarel F, Borde S, Genova F,
  Jasniewicz G, Lalo{\"e} S, Lesteven S et~al. (2000) The simbad astronomical
  database-the cds reference database for astronomical objects.
\newblock \emph{Astronomy and Astrophysics Supplement Series} 143(1): 9--22.

\bibitem[{Westphal et~al.(2005)Westphal, Butterworth, Snead, Craig, Anderson,
  Jones, Brownlee, Farnsworth and Zolensky}]{westphal2005stardust}
Westphal AJ, Butterworth AL, Snead CJ, Craig N, Anderson D, Jones SM, Brownlee
  DE, Farnsworth R and Zolensky ME (2005) Stardust@ home: a massively
  distributed public search for interstellar dust in the stardust interstellar
  dust collector.
\newblock \emph{Lunar and Planetary Science XXXVI, Part 21} .

\bibitem[{Wheeler and Kipping(2019)}]{wheeler2019weird}
Wheeler A and Kipping D (2019) The weird detector: flagging periodic, coherent
  signals of arbitrary shape in time-series photometry.
\newblock \emph{Monthly Notices of the Royal Astronomical Society} 485(4):
  5498--5510.

\bibitem[{Winters et~al.(2022)Winters, Cloutier, Medina, Irwin, Charbonneau,
  Astudillo-Defru, Bonfils, Howard, Isaacson, Bean et~al.}]{winters2022second}
Winters JG, Cloutier R, Medina AA, Irwin JM, Charbonneau D, Astudillo-Defru N,
  Bonfils X, Howard AW, Isaacson H, Bean JL et~al. (2022) A second planet
  transiting ltt 1445a and a determination of the masses of both worlds.
\newblock \emph{The Astronomical Journal} 163(4): 168.

\bibitem[{Wright et~al.(2010)Wright, Eisenhardt, Mainzer, Ressler, Cutri,
  Jarrett, Kirkpatrick, Padgett, McMillan, Skrutskie et~al.}]{wright2010wide}
Wright EL, Eisenhardt PR, Mainzer AK, Ressler ME, Cutri RM, Jarrett T,
  Kirkpatrick JD, Padgett D, McMillan RS, Skrutskie M et~al. (2010) The
  wide-field infrared survey explorer (wise): mission description and initial
  on-orbit performance.
\newblock \emph{The Astronomical Journal} 140(6): 1868.

\bibitem[{Wright et~al.(2015)Wright, Cartier, Zhao, Jontof-Hutter and
  Ford}]{wright2015g}
Wright JT, Cartier KM, Zhao M, Jontof-Hutter D and Ford EB (2015) The {\^g}
  search for extraterrestrial civilizations with large energy supplies. iv. the
  signatures and information content of transiting megastructures.
\newblock \emph{The Astrophysical Journal} 816(1): 17.

\bibitem[{Yu et~al.(2019)Yu, Thorstensen, Rappaport, Mann, Jacobs, Nelson,
  Gaensicke, LaCourse, Borkovits, Aiken et~al.}]{yu20199}
Yu Z, Thorstensen J, Rappaport S, Mann A, Jacobs T, Nelson L, Gaensicke BT,
  LaCourse D, Borkovits T, Aiken J et~al. (2019) A 9-hr cv with one outburst in
  4 years of kepler data.
\newblock \emph{Monthly Notices of the Royal Astronomical Society} .

\bibitem[{Zhou et~al.(2018)Zhou, Rappaport, Nelson, Huang, Senhadji, Rodriguez,
  Vanderburg, Quinn, Johnson, Latham et~al.}]{zhou2018occultations}
Zhou G, Rappaport S, Nelson L, Huang C, Senhadji A, Rodriguez J, Vanderburg A,
  Quinn S, Johnson C, Latham D et~al. (2018) Occultations from an active
  accretion disk in a 72-day detached post-algol system detected by k2.
\newblock \emph{The Astrophysical Journal} 854(2): 109.

\bibitem[{Zink et~al.(2019)Zink, Hardegree-Ullman, Christiansen, Crossfield,
  Petigura, Lintott, Livingston, Ciardi, Barentsen, Dressing
  et~al.}]{zink2019catalog}
Zink JK, Hardegree-Ullman KK, Christiansen JL, Crossfield IJ, Petigura EA,
  Lintott CJ, Livingston JH, Ciardi DR, Barentsen G, Dressing CD et~al. (2019)
  Catalog of new k2 exoplanet candidates from citizen scientists.
\newblock \emph{Research Notes of the AAS} 3(2): 43.

\end{thebibliography}

\end{document}